\documentclass[12pt]{article}

\usepackage{amsmath,amssymb,array,multirow,amsfonts}
\usepackage{bm}
\usepackage{tabstackengine}
\usepackage{graphicx}
\usepackage{hyperref}
\usepackage{authblk}

\title{Bayesian estimation of a multivariate TAR model when the noise process distribution belongs to the class of Gaussian variance mixtures} 

\author[1]{L.H. Vanegas}
\author[1]{S.A. Calderón }
\author[1]{L.M. Rondón }

\affil[1]{Universidad Nacional de Colombia, Sede Bogotá}

\begin{document}
\maketitle



%
%
                          


\begin{abstract}
A threshold autoregressive (TAR) model is a powerful tool for analyzing nonlinear multivariate time series, which includes special cases like self-exciting threshold autoregressive (SETAR) models and vector autoregressive (VAR) models. In this paper, estimation, inference, and forecasting using the Bayesian approach are developed for multivariate TAR (MTAR) models considering a flexible setup, under which the noise process behavior can be described using not only the Gaussian distribution but also other distributions that belong to the class of Gaussian variance mixtures, which includes Student-$t$, Slash, symmetric hyperbolic, and contaminated normal distributions, which are also symmetric but are more flexible and with heavier tails than the Gaussian one. Inferences from MTAR models based on that kind of distribution may be less affected by extreme or outlying observations than those based on the Gaussian one. All parameters in the MTAR model are included in the proposed MCMC-type algorithm, except the number of regimes and the autoregressive orders, which can be chosen using the Deviance Information Criterion (DIC) and/or the Watanabe-Akaike Information Criterion (WAIC). A library for the language and environment for statistical computing {\tt R} was also developed to assess the effectiveness of the proposed methodology using simulation studies and analysis of two real multivariate time series.
\end{abstract}






\section{Introduction}
\label{sec1}
Time series analysis is one of the most interesting topics in Data Science due to its applicability to various fields, such as hydrology, economics, engineering, and epidemiology, among others. Due to the importance of understanding how variables of interest will behave in the future, statistics and machine learning are currently addressing modeling and forecasting in time series. To accomplish these tasks, linear and non-linear statistical models, neural network algorithms, and generative models have been developed for modeling and predicting univariate and multivariate time series. 

In \cite{Tsay1998}, a multivariate TAR (MTAR) model is introduced along with an estimation strategy as one of the first works dealing with multivariate and non-linear time series. A review of thresholds and smooth transition vector autoregressive models can also be found in \cite{Hubrich2023}. Further, GARCH models can be used to analyze multivariate time series that present volatility, whose non-linearity is explained by the equation of conditional variance; see \cite{Wei2019}, \cite{Tsay2013}, \cite{Lut2005}. Nonlinear and non-Gaussian state space models are flexible models that can be used to analyze and predict univariate and multivariate time series in many contexts; see \cite{Triantafyllopoulos2021} and \cite{TANIZAKI2003871}. While all methods presented above were based on a parametric approach, there are also proposals for a non-parametric approach to analyzing multivariate time series. See \cite{HARDLE1998221}, \cite{Jeliazkov2013}, \cite{KALLI2018267}, and \cite{Samadi2019}.

The literature above and others not included here assume that innovations can be adequately described using the Gaussian distribution, especially for MTAR models. Even so, the Gaussian assumption is only appropriate for some applications. As an example, financial time series tend to follow a distribution with heavy tails. Some works have attempted to introduce these kinds of distributions into MTAR models. For example, \cite{RC21} presented a Bayesian procedure for estimating non-structural parameters when innovations follow a multivariate Student-$t$ distribution. \cite{Abril2018} describes a Bayesian approach based on adaptive Monte Carlo Markov Chain (MCMC) to estimate autoregressive coefficients and covariance matrices in MTAR models when the noise process behaves according to the generalized error distribution. However, that methodology does not provide a procedure to forecast and identify non-structural parameters. The purpose of this paper is to present a Bayesian approach for estimating non-structural parameters and forecasting MTARs when the noise process distribution belongs to the class of Gaussian variance mixtures that includes Student-$t$, Slash, symmetric hyperbolic, and contaminated normal distributions, which are also symmetric but are more flexible and with heavier tails than the Gaussian one. The number of regimes and the autoregressive orders can be chosen using the Deviance Information Criterion (DIC) and/or the Watanabe-Akaike Information Criterion (WAIC).

The rest of this paper is organized as follows. First, we present the theoretical framework on which the MTAR model and the family of multivariate Gaussian variance mixtures are based. In addition, the Bayesian methodology for estimating non-structural parameters and threshold values via MCMC methods is discussed in that section. Afterward, we introduce the procedure for sampling from the predictive distribution used in forecasting. After this, a simulation experiment is conducted to assess the effectiveness of the estimation and forecasting methodology. To illustrate the proposed methodology, a real-data application is presented in the context of Colombian river flows and financial data. Lastly, conclusions and recommendations are offered.

\section{Theoretical Framework}
\label{sec2}

\subsection{The MTAR Model}
\label{subsec1}
Let $\{Y_t\}$ and $\{X_t\}$ be multivariate stochastic processes such that $Y_t=(Y_{1t},\ldots,Y_{kt})^{\!\top}$ and $X_t=(X_{1t},\ldots,X_{rt})^{\!\top}$,
and let $\{Z_t\}$ be an univariate process. According to \cite{CN17} and \cite{RC21}, the output stochastic process $\{Y_t\}$ is said to follow a Multivariate Threshold Auto Regressive (MTAR) model with co-variate vector $X_t$ and threshold variable $Z_t$, denoted here by ${\rm MTAR}(l;p,q,d)$, where $p=(p_1,\ldots,p_l)^{\!\top}$, $q=(q_1,\ldots,q_l)^{\!\top}$ and $d=(d_1,\ldots,d_l)^{\!\top}$ are vectors of non-negative integer values, if
\begin{equation*}
	\resizebox{\columnwidth}{!}{$Y_t={\phi}_0^{^{(j)}}+\sum\limits_{i=1}^{p_j}\bm{\phi}_i^{^{(j)}}Y_{t-i}+\sum\limits_{i=1}^{q_j}\bm{\beta}_i^{^{(j)}}X_{t-i}+\sum\limits_{i=1}^{d_j}{\delta}_i^{^{(j)}}Z_{t-i}+\bm{\Sigma}_j^{\!^{\frac{1}{2}}}\epsilon_t\qquad\text{when}\quad Z_{t-h}\in(c_{j-1},c_j]$}
\end{equation*}

for some $j\in\{1,\ldots,l\}$. The real numbers $c_1< c_2 < \cdots < c_{l-1}$, which form the parameter vector ${c}=(c_1, c_2, \cdots,c_{l-1})^{\!\top}$, are referred to as threshold values of the process $\{Z_t\}$, and because $c_0=-\infty$ and $c_l=\infty$, they define $l$ regimes for the process $\{Z_t\}$. In the regime $j$-th, the dimensions of the location parameters ${\phi}_0^{^{(j)}}$, $\bm{\phi}_i^{^{(j)}}$, $\bm{\beta}_i^{^{(j)}}$ and ${\delta}_i^{^{(j)}}$ are $k\times 1$, $k\times k$, $k\times r$ and $k\times 1$, respectively, while the scale parameter $\bm{\Sigma}_j$ is a $k\times k$ positive definite matrix. In the regime $j$-th, the value of $p_j$ is the autoregressive order, while $q_j$ and $d_j$ are the maximum lags for the covariate vector $X_t$ and the threshold variable $Z_t$, respectively. Moreover, $h$ represents a non-negative integer value commonly referred to as the delay parameter, and $\{\epsilon_t\}$ is a white noise process, which is mutually independent of $\{X_t\}$ and $\{Z_t\}$. The values of $l$, $p$, $q$ and $d$ are assumed to be known.

Special cases of the above setup include self-exciting threshold autoregressive models (SETAR, see \cite{TL80}) when $k=1$, $Z_t=Y_t$ for all $t$, and $h>0$; and vector autoregressive models (VAR, see \cite{L07}) when $l=1$. In the $j$-th regime, the MTAR model can be expressed as the following multivariate multiple linear model:
\begin{equation}
	\bm{Y}_j=\bm{M}_j\bm{\theta}_j+\bm{\epsilon}_j, 
	\label{LMMTAR}
\end{equation}

where
\begin{itemize}
	\item $\bm{Y}_j=(Y_{t^{^{(j)}}_1},\ldots,Y_{t^{^{(j)}}_{n_j}})^{\!\top}$ is the $n_j\times k$ response matrix, in which $t^{^{(j)}}_1,\ldots,t^{^{(j)}}_{n_j}$ represent the time points $t$ such that $Z_{t-h}\in(c_{j-1},c_j]$ for 
	
	$t>{\rm max}\{p_1,\ldots,p_l,q_1,\ldots,q_l,d_1,\ldots,d_l,h\}$.
	\item \resizebox{\columnwidth}{!}{$\bm{M}_j
				=\begin{bmatrix}
					M_{t^{^{(j)}}_1}\\ \vdots\\ M_{t^{^{(j)}}_{n_j}}
				\end{bmatrix}
				=\begin{bmatrix}
					1      & y_{t^{^{(j)}}_1-1}^{\!\top} & \ldots & y_{t^{^{(j)}}_1-p_j}^{\!\top} & x_{t^{^{(j)}}_1-1}^{\!\top} & \ldots & x_{t^{^{(j)}}_1-q_j}^{\!\top} & z_{t^{^{(j)}}_1-1} & \ldots & z_{t^{^{(j)}}_1-d_j}\\
					\vdots & \vdots           & \vdots & \vdots             & \vdots           & \vdots & \vdots             & \vdots  & \vdots & \vdots  \\
					1      & y_{t^{^{(j)}}_{n_j}-1}^{\!\top} & \ldots & y_{t^{^{(j)}}_{n_j}-p_j}^{\!\top} & x_{t^{^{(j)}}_{n_j}-1}^{\!\top} & \ldots & x_{t^{^{(j)}}_{n_j}-q_j}^{\!\top} & z_{t^{^{(j)}}_{n_j}-1} & \ldots & z_{t^{^{(j)}}_{n_j}-d_j}
				\end{bmatrix}$}

	is the known $n_j\times s_j$ design/model matrix, in which $s_j=1 + (p_j\times k) + (q_j\times r) + d_j$, and $y_{t^{^{(j)}}_i}$, $x_{t^{^{(j)}}_i}$ and $z_{t^{^{(j)}}_i}$  represent, respectively, the realizations of $Y_{t^{^{(j)}}_i}$, $X_{t^{^{(j)}}_i}$ and $Z_{t^{^{(j)}}_i}$.
	\item $\bm{\theta}_j=({\phi}_0^{^{(j)}},\bm{\phi}_1^{^{(j)}},\ldots,\bm{\phi}_{p_j}^{^{(j)}},\bm{\beta}_1^{^{(j)}},\ldots,\bm{\beta}_{q_j}^{^{(j)}},{\delta}_1^{^{(j)}},\ldots,{\delta}_{d_j}^{^{(j)}})^{\!\top}$ is the $s_j\times k$ unknown regression parameter matrix.
	\item $\bm{\epsilon}_j=(\bm{\Sigma}_j^{\!^{\frac{1}{2}}}\epsilon_{t^{^{(j)}}_{1}},\ldots,\bm{\Sigma}_j^{\!^{\frac{1}{2}}}\epsilon_{t^{^{(j)}}_{n_j}})^{\!\top}$ is the $n_j\times k$ random error matrix. The rows of $\bm{\epsilon}_j$ are independent and identically distributed random vectors with location and scale parameters given, respectively, by $0$ and $\bm{{\Sigma}}_j$, and whose distribution is assumed to belong to the class of Gaussian variance mixtures.
\end{itemize}

\noindent If $\epsilon_t$ is assumed to follow a multivariate Gaussian distribution, then $\bm{\epsilon}_j\sim {\rm MN}_{n_j,k}(\bm{M}_j\bm{\theta}_j,\bm{I}_{n_j},\bm{\Sigma}_j)$, that is, $\bm{\epsilon}_j$ follows a matrix Gaussian distribution (see, for instance, \cite[chapter 2]{GN99}) with location parameter $\bm{M}_j\bm{\theta}_j$ and scale parameters $\bm{I}_{n_j}$ and $\bm{\Sigma}_j$, in which $\bm{I}_{n_j}$ represents the 
identity matrix of order $n_j$. Therefore, given the values of $h$ and $c$, the likelihood function of $\alpha=({\rm vec}(\bm{\theta}_1)^{\!\top},{\rm vec}(\bm{\Sigma}_1)^{\!\top},\ldots,{\rm vec}(\bm{\theta}_l)^{\!\top},{\rm vec}(\bm{\Sigma}_l)^{\!\top})^{\!\top}$ is the following:
$$
L(\alpha)=\prod\limits_{j=1}^l \dfrac{\exp\!\left(\!-\frac{1}{2}{\rm tr}\!\!\left[(\bm{y}_j-\bm{M}_j\bm{\theta}_j)^{\!\top}\!(\bm{y}_j-\bm{M}_j\bm{\theta}_j)\bm{\Sigma}_j^{-1}\right]\!\right)}{(2\pi)^{\frac{n_j\,k}{2}}\,|\bm{\Sigma}_j|^{\frac{n_j}{2}}},
$$
where $\bm{y}_j$ represents the realization of $\bm{Y}_j$.
Thus, the maximum likelihood estimators of $\bm{\theta}_j$ and $\bm{\Sigma}_j$ are given by
$$\hat{\bm{\theta}}_j=(\bm{M}_j^{\!\top}\bm{M}_j)^{\!-1}\bm{M}_j^{\!\top}\bm{Y}_j\qquad\quad\text{and}\qquad \hat{\bm{\Sigma}}_j=\dfrac{1}{n_j}\!(\bm{Y}_j-\bm{M}_j\hat{\bm{\theta}}_j)^{\!\top}(\bm{Y}_j-\bm{M}_j\hat{\bm{\theta}}_j).$$
Therefore, $\hat{\bm{\theta}}_j\sim {\rm MN}_{s_j,k}(\bm{\theta}_j,(\bm{M}_j^{\!\top}\bm{M}_j)^{\!-1},\bm{\Sigma}_j)$ and
${\rm vec}(\hat{\bm{\theta}}_j)\sim {\rm Normal}_{s_j\times k}({\rm vec}(\bm{\theta}_j),\bm{\Sigma}_j\!\otimes\!(\bm{M}_j^{\!\top}\bm{M}_j)^{\!-1})$, where
$\otimes$ represents the Kronecker product.

\subsection{Multivariate Gaussian variance mixtures}
\label{subsec22}
The distribution of the $k$-dimensional random vector $Y$ belongs to the class of multivariate Gaussian variance mixtures (\cite{AM74}; \cite[section 6.2]{MFE05}) if it can be written as $Y=\mu+\sqrt{\kappa(U)}\epsilon_0$, where $\kappa(\cdot)$ is a positive function, $\epsilon_0\sim {\rm Normal}_k(0,\bm{\Sigma})$, $\mu=(\mu_1,\ldots,\mu_k)^{\!\top}$, $\bm{\Sigma}$ is a positive definite matrix and $U$ is a random variable independent of $\epsilon_0$ and whose probability density function, denoted here by $f_U(u|\nu)$, 
may be dependent on an {\it extra parameter} denoted here by $\nu$, in which ${\rm Normal}_k(\mu,\bm{\Sigma})$ denotes the $k$-dimensional multivariate Gaussian distribution with location and scale parameters given, respectively, by $\mu$ and $\bm{\Sigma}$. Therefore, $Y|U=u \sim {\rm Normal}_k(\mu,\kappa(u)\bm{\Sigma})$, and the probability density function of $Y$ can be obtained as follows

 \resizebox{\columnwidth}{!}{
$
f_{Y}(y|\mu,\bm{\Sigma},\nu)=
\begin{cases}
	\int\dfrac{1}{(2\pi)^{\frac{k}{2}}|\kappa(u)\bm{\Sigma}|^{\frac{1}{2}}}\exp\!\!\left(\!-\frac{1}{2}(y-\mu)^{\!\top}\![\kappa(u)\bm{\Sigma}]^{-1}\!(y-\mu)\!\right)\!f_U(u|\nu) du &\text{if }U\text{ is continuous}\\
	\sum\limits_{u}\dfrac{1}{(2\pi)^{\frac{k}{2}}|\kappa(u)\bm{\Sigma}|^{\frac{1}{2}}}\exp\!\!\left(\!-\frac{1}{2}(y-\mu)^{\!\top}\![\kappa(u)\bm{\Sigma}]^{-1}\!(y-\mu)\!\right)\!f_U(u|\nu) &\text{if }U\text{ is discrete}
\end{cases}
$}
If ${\rm E}(Y)$ and ${\rm Var}(Y)$ exist, then they are given, respectively, by $\mu$ and ${\rm E}[\kappa(U)]\bm{\Sigma}$. In addition, if the $k$-dimensional random vector $\epsilon$ belongs to the class of multivariate Gaussian variance mixtures with location parameter $0$, scale parameter ${\bf I}_k$, and extra parameter $\nu$, then $Y=\mu + \bm{\Sigma}^{\frac{1}{2}}\epsilon$, where $\mu=(\mu_1,\ldots,\mu_k)^{\!\top}$ and $\bm{\Sigma}$ is a $k\times k$ positive 
definite matrix, also belongs to the class of multivariate Gaussian variance mixtures with location parameter $\mu$, scale parameter $\bm{\Sigma}$, and extra parameter $\nu$ (\cite[proposition 6.9]{MFE05}). Table \ref{smgd} lists some distributions that belong to the class of multivariate Gaussian variance mixtures. Many of these distributions are more flexible than the Gaussian distribution, and some of them also have heavier tails. As a consequence, inferences from MTAR models based on those distributions may be less influenced by extreme or outlying observations than inferences from models based on the Gaussian distribution.

\subsubsection{Multivariate Student-$t$ distribution}

\label{subsubsec31}
This distribution is obtained by using $U\sim {\rm Gamma}(\frac{\nu}{2},\frac{\nu}{2})$ and $\kappa(u)=u^{-1}$. Thus, the probability density function of $Y$ becomes (\cite[example 6.7]{MFE05}; \cite[page 85]{FKN18})
\begin{align*}
	f_{Y}(y|\mu,\bm{\Sigma},\nu)&=\dfrac{\left(\frac{\nu}{2}\right)^{\!\frac{\nu}{2}}}{\Gamma\!\left(\frac{\nu}{2}\right)(2\pi)^{\frac{k}{2}}|\bm{\Sigma}|^{\frac{1}{2}}}\int\limits_{0}^{\infty}\underbrace{u^{\frac{k+\nu}{2}-1}\exp\!\left(\!-\frac{u}{2}\!\!\left[\nu+(y-\mu)^{\!\top}\bm{\Sigma}^{\!-1}(y-\mu)\right]\!\right)}_{\text{kernel of Gamma distribution}}d u\\
	&=\dfrac{\Gamma(\frac{\nu+k}{2})}{\Gamma(\frac{\nu}{2})(\nu\pi)^{\frac{k}{2}}|\bm{\Sigma}|^{\frac{1}{2}}}\!\left(\!\!1+\frac{1}{\nu}(y-\mu)^{\!\top}\bm{\Sigma}^{\!-1}(y-\mu)\!\!\right)^{\!\!-\frac{\nu+k}{2}},\qquad \nu>0.
\end{align*}
In addition, ${\rm E}(Y)=\mu$ for $\nu>1$ and ${\rm Var}(Y)=\left(\!\dfrac{\nu}{\nu-2}\!\right)\!\bm{\Sigma}$ for $\nu>2$.
The distribution of $Y$ has heavier tails than that of the Gaussian one. The ${\rm Normal}_k(\mu,\bm{\Sigma})$ is a limiting case when $\nu\to\infty$, 
because $U^{-1}$ converges in probability to $1$ as $\nu\to\infty$.

\subsubsection{Multivariate Slash distribution}
\label{subsubsec41}
This distribution is obtained by using $U\sim {\rm Beta}(\frac{\nu}{2},1)$ and $\kappa(u)=u^{-1}$. Thus,
\begin{align*}
	f_{Y}(y|\mu,\bm{\Sigma},\nu)&=\dfrac{\frac{\nu}{2}}{(2\pi)^{\frac{k}{2}}|\bm{\Sigma}|^{\frac{1}{2}}}\int\limits_{0}^1 u^{\frac{k+\nu}{2}-1}\exp\!\left(\!-\frac{u}{2}(y-\mu)^{\!\top}\bm{\Sigma}^{\!-1}(y-\mu)\!\right)d u,\qquad \nu>0.
\end{align*}
Therefore, the probability density function of $Y$ becomes \cite{WG2006,MAAI17}
$$
f_{Y}(y|\mu,\bm{\Sigma},\nu)=
\begin{cases}
	\dfrac{\nu}{(2\pi)^{\frac{k}{2}}|\bm{\Sigma}|^{\frac{1}{2}}(\nu+k)} \qquad &\text{if}\quad y=\mu\\
	\\
	\dfrac{\frac{\nu}{2}\gamma(\frac{k+\nu}{2},\frac{1}{2}(y-\mu)^{\!\top}\bm{\Sigma}^{\!-1}(y-\mu))}{(2\pi)^{\frac{k}{2}}|\bm{\Sigma}|^{\frac{1}{2}}\!\left[\frac{1}{2}(y-\mu)^{\!\top}\bm{\Sigma}^{\!-1}(y-\mu)\right]^{\frac{k+\nu}{2}}} \qquad &\text{if}\quad y\neq\mu, \\
\end{cases}
$$
where $\gamma(a,b)=\int\limits_{0}^b t^{a-1}\exp(-t)dt$ represents the lower incomplete gamma function. In addition, ${\rm E}(Y)=\mu$ for $\nu>1$ and ${\rm Var}(Y)=\left(\!\dfrac{\nu}{\nu-2}\!\right)\!\bm{\Sigma}$ for $\nu>2$. The distribution of $Y$ has heavier tails than that of the Gaussian one. The ${\rm Normal}_k(\mu,\bm{\Sigma})$ is a limiting case when $\nu\to\infty$, 
because $U^{-1}$ converges in probability to $1$ as $\nu\to\infty$.

\subsubsection{Multivariate contaminated Gaussian distribution}
\label{subsubsec51}
This distribution is obtained by using $U\sim 1 - (1-\nu_2){\rm Bernoulli}(\nu_1)$ (that is, $U=\nu_2$ with probability $\nu_1$ and $U=1$ with probability $1-\nu_1$) and $\kappa(u)=u^{-1}$. Thus, the probability density function of $Y$ becomes \cite{Tukey60}
\begin{scriptsize}

\begin{align*}
	f_{Y}(y|\mu,\bm{\Sigma},\nu_1,\nu_2)&={\rm Pr}(U=\nu_2){\rm Normal}(\mu,u_2^{-1}\bm{\Sigma}) + {\rm Pr}(U=1){\rm Normal}(\mu,\bm{\Sigma})\\ 
	&=\nu_1\dfrac{u_2^{\frac{k}{2}}}{(2\pi)^{\frac{k}{2}}|\bm{\Sigma}|^{\frac{1}{2}}}\exp\!\left(\!-\frac{u_2}{2}(y-\mu)^{\!\top}\bm{\Sigma}^{\!-1}(y-\mu)\!\right) + (1-\nu_1)\dfrac{\exp\!\left(\!-\frac{1}{2}(y-\mu)^{\!\top}\bm{\Sigma}^{\!-1}(y-\mu)\!\right)}{(2\pi)^{\frac{k}{2}}|\bm{\Sigma}|^{\frac{1}{2}}},
\end{align*}

\end{scriptsize}
where $\nu_1,\nu_2\in (0,1)$. The distribution of $Y$ is a special case ($d=k$, $p=\nu_1$, $k_1=\nu_2$ and $k_2=1$) of the multivariate two-point Gaussian mixture distribution described in \cite[example 6.6]{MFE05}.
In addition, ${\rm E}(Y)=\mu$ and ${\rm Var}(Y)=\left(\!\dfrac{\nu_1}{\nu_2}+1-\nu_1\!\!\right)\!\!\bm{\Sigma}$.
The distribution of $Y$ has heavier tails than those of the Gaussian one. The ${\rm Normal}_k(\mu,\bm{\Sigma})$ is a limiting case when $\nu_1\to 0$ or $\nu_2\to 1$, because $U^{-1}$ converges in probability to $1$ as $\nu_1\to 0$ or $\nu_2\to 1$.

\subsubsection{Multivariate symmetric Hyperbolic distribution}
\label{subsubsec61}
This distribution is obtained by using $U\sim {\rm GIG}(1,1,\nu^2)$ and $\kappa(u)=u$. Thus, the probability density function of $Y$ becomes 
\begin{footnotesize}
\begin{align*}
	f_{Y}(y|\mu,\bm{\Sigma},\nu)&=\dfrac{\nu}{2K_{1}(\nu)(2\pi)^{\frac{k}{2}}|\bm{\Sigma}|^{\frac{1}{2}}}\int\limits_{0}^{\infty}\underbrace{u^{\frac{2-k}{2}-1}\exp\!\left(\!-\frac{1}{2}\!\left[\!\frac{1}{u}\!\left(1 + (y-\mu)^{\!\top}\bm{\Sigma}^{\!-1}(y-\mu)\!\right)+u\nu^2\right]\!\right)}_{\text{kernel of Generalized Inverse Gaussian distribution}}d u\\
	&=\dfrac{\nu^{\frac{k}{2}} K_{\frac{2-k}{2}}\!\!\left(\!\nu\sqrt{1 + (y-\mu)^{\!\top}\bm{\Sigma}^{\!-1}(y-\mu)}\right) \left(\!\!\sqrt{1+(y-\mu)^{\!\top}\bm{\Sigma}^{\!-1}(y-\mu)}\right)^{\!\!\frac{2-k}{2}}}{K_{1}(\nu)(2\pi)^{\frac{k}{2}}|\bm{\Sigma}|^{\frac{1}{2}}},
\end{align*}
\end{footnotesize}
where $K_{a}(b)=\int\limits_{0}^{\infty}x^{a-1}\exp\!\!\left(\!-\frac{b}{2}(x+x^{-1})\!\right)d x$ is the modified
Bessel function of third-order and index $a$ (see, for example, \cite[section 9.6]{AS65}).
The distribution of $Y$ is a special case ($d=k$, $\lambda=1$, $\chi=1$ and $\psi=\nu^2$) of the multivariate symmetric generalized hyperbolic distribution described in \cite[example 6.8]{MFE05}. Furthermore, ${\rm E}(Y)=\mu$ and ${\rm Var}(Y)=\left(\!\dfrac{K_2(\nu)}{\nu K_1(\nu)}\!\right)\!\!\bm{\Sigma}$.
The distribution of $Y$ has heavier (lighter) tails than those of the Gaussian one for ``small'' (``large'') values of $\nu$.

\subsubsection{Multivariate Laplace distribution}
\label{subsubsec71}
This distribution is obtained using $U\sim {\rm Exponential}(\frac{1}{8})$ and $\kappa(u)=u$. Thus, the probability density function of $Y$ becomes \cite[page 234]{KKP01}

\begin{scriptsize}

\begin{align*}
	f_{Y}(y|\mu,\bm{\Sigma})&=\dfrac{1}{8(2\pi)^{\frac{k}{2}}|\bm{\Sigma}|^{\frac{1}{2}}}\int\limits_{0}^{\infty}\underbrace{u^{\frac{2-k}{2}-1}\exp\!\left(\!-\frac{1}{2}\!\!\left[\frac{1}{u}(y-\mu)^{\!\top}\bm{\Sigma}^{\!-1}(y-\mu)+\frac{u}{4}\right]\!\right)}_{\text{kernel of Generalized Inverse Gaussian distribution if }y\neq\mu}d u\\
	&=\dfrac{1}{2^{k+1}\pi^{\frac{k}{2}}|\bm{\Sigma}|^{\frac{1}{2}}}\!
	\left(\!\sqrt{(y-\mu)^{\!\top}\bm{\Sigma}^{\!-1}(y-\mu)}\right)^{\!\!\frac{2-k}{2}}\!
	K_{\frac{2-k}{2}}\!\!\left(\!\frac{1}{2}\sqrt{(y-\mu)^{\!\top}\bm{\Sigma}^{\!-1}(y-\mu)}\right),\qquad y\neq \mu.
\end{align*}
\end{scriptsize}

In addition, ${\rm E}(Y)=\mu$ and ${\rm Var}(Y)=8\bm{\Sigma}$. The distribution of $Y$ has heavier tails than those of the Gaussian one.

\begin{table}[!ht]
	\centering
	\caption{Some Gaussian variance mixture distributions.}\label{smgd}
\resizebox{\columnwidth}{!}{
		\begin{tabular}{c|c|c|c}\hline                                         
			\multirow{2}{*}{Distribution} & \multirow{2}{*}{$\kappa(u)$} &   \multicolumn{2}{c}{Mixing distribution} \\\cline{3-4}
			&               &    Family         &           $f_U(u|\nu)$ \\\hline
			&               &                   & \multirow{3}{*}{$\dfrac{\left(\frac{\nu}{2}\right)^{\!\frac{\nu}{2}}}{\Gamma\!\left(\frac{\nu}{2}\right)}\,u^{\frac{\nu}{2}-1}\!\exp\!\left(\!-\frac{\nu}{2}u\!\right)\!\!I_{(0,\infty)}(u),\quad\nu>0$}\\
			{Student-$t(\nu)$}      & {$u^{-1}$}       &{Gamma}$\left(\frac{\nu}{2},\frac{\nu}{2}\right)$         &                                                                                                                                                                         \\
			&               &                   &                                                                                                                                                                         \\\hline 
			&               &                   & \multirow{3}{*}{$\frac{\nu}{2} u^{\frac{\nu}{2}-1}\!I_{(0,1)}(u),\quad\nu>2$} \\
			Slash$(\nu)$             &        $u^{-1}$  & Beta$(\frac{\nu}{2},1)$      &                                  \\
			&               &                   &                                  \\\hline
			Contaminated      &               & \multirow{3}{*}{$1 - (1-\nu_2){\rm Bernoulli}(\nu_1)$}         & \multirow{3}{*}{$\begin{cases} \nu_1   &\mbox{if }u=\nu_2\\ 1-\nu_1 &\mbox{if }u=1 \end{cases},\quad \nu_1,\nu_2\in(0,1)$} \\
			normal            &        $u^{-1}$  &                   & \\
			$(\nu_1,\nu_2)$                  &               &                   & \\\hline
			Symmetric         &               & Generalized       & \multirow{3}{*}{$\dfrac{\nu}{2\,{\rm K}_{1}(\nu)}\exp\!\left(-\frac{1}{2}\!\left(\frac{1}{u}+\nu^2 u\right)\right)\!\!I_{(0,\infty)}(u),\quad\nu>0$}\\
			hyperbolic        &        $u$    & Inverse Gaussian  & \\
			$(\nu)$                 &               &   $(1,1,\nu^2)$   & \\\hline
			Laplace or        &               &                   & \multirow{3}{*}{$\dfrac{1}{8}\exp\!\left(-\frac{1}{8}u\right)\!\!I_{(0,\infty)}(u)$}            \\
			double            &        $u$    & Exponential$(\frac{1}{8})$       & \\
			exponential       &               &                   & \\\hline
	\end{tabular}} 
\end{table}

\subsection{Augmented likelihood function}
\label{subsec23}
If the realization of the random vector $U_j=(U_{t^{^{(j)}}_{1}},\ldots,U_{t^{^{(j)}}_{n_j}})^{\!\top}$, denoted here by $u_j=(u_{t^{^{(j)}}_{1}},\ldots,u_{t^{^{(j)}}_{n_j}})^{\!\top}$, is in fact observed, then 
$$\bm{\epsilon}_j\sim {\rm MN}_{n_j,k}(\bm{0},\bm{\kappa}(u_j),\bm{\Sigma}_j)\implies \bm{Y}_j\sim {\rm MN}_{n_j,k}(\bm{M}_j\bm{\theta}_j,\bm{\kappa}(u_j),\bm{\Sigma}_j),$$
where $\bm{\kappa}(u_j)\equiv {\rm diag}\{\kappa(u_{t^{^{(j)}}_{1}}),\ldots,\kappa(u_{t^{^{(j)}}_{n_j}})\}$. Therefore, the augmented likelihood function of the vector of the interest parameter $\alpha$ (that is, the likelihood function obtained by assuming that the realizations of the random vectors $U_j$ for $j=1,\ldots,l$ are actually observed) can be expressed as follows
\begin{scriptsize}

\begin{align*}
	L(\alpha)&=\prod\limits_{j=1}^l \dfrac{\exp\!\left(\!-\frac{1}{2}{\rm tr}\!\!\left[(\bm{y}_j-\bm{M}_j\bm{\theta}_j)^{\!\top}\![\bm{\kappa}(u_j)]^{-1}\!(\bm{y}_j-\bm{M}_j\bm{\theta}_j)\bm{\Sigma}_j^{-1}\right]\!\right)}{(2\pi)^{\frac{n_j\,k}{2}}|\bm{\kappa}(u_j)|^{\frac{k}{2}}\,\,|\bm{\Sigma}_j|^{\frac{n_j}{2}}}\\
	&=\prod\limits_{j=1}^l \dfrac{\exp\!\left(\!-\frac{1}{2}{\rm tr}\!\left\{\!\left[(\bm{y}_j-\bm{M}_j\widetilde{\bm{\theta}}_j)^{\!\top}\![\bm{\kappa}(u_j)]^{-1}\!(\bm{y}_j-\bm{M}_j\widetilde{\bm{\theta}}_j)\! +\! (\bm{\theta}_j-\widetilde{\bm{\theta}}_j)^{\!\top}\!(\bm{M}_j^{\!\top}\![\bm{\kappa}(u_j)]^{-1}\!\bm{M}_j)(\bm{\theta}_j-\widetilde{\bm{\theta}}_j)\right]\!\bm{\Sigma}_j^{-1}\!\right\}\right)}{(2\pi)^{\frac{n_j\,k}{2}}|\bm{\kappa}(u_j)|^{\frac{k}{2}}\,\,|\bm{\Sigma}_j|^{\frac{n_j}{2}}},
\end{align*}
\end{scriptsize}
where $\widetilde{\bm{\theta}}_j=(\bm{M}_j^{\!\top}\![\bm{\kappa}(u_j)]^{-1}\!\bm{M}_j)^{-1}\bm{M}_j^{\!\top}\![\bm{\kappa}(u_j)]^{-1}\bm{y}_j$. By following the properties of the matrix Gaussian distribution, it is also possible to express the likelihood function of $\alpha$ as follows:
$$L(\alpha)=\prod\limits_{j=1}^l \prod\limits_{i=1}^{n_j} \dfrac{\exp\!\left(\!\!-\frac{1}{2}({y}_{t^{^{(j)}}_{i}}^{\!\top}-{M}_{t^{^{(j)}}_{i}}\bm{\theta}_j)[\kappa(u_{t^{^{(j)}}_{i}})\bm{\Sigma}_j]^{-1}\!({y}_{t^{^{(j)}}_{i}}^{\!\top}-{M}_{t^{^{(j)}}_{i}}\bm{\theta}_j)^{\!\top}\!\right)}{(2\pi)^{\frac{k}{2}}\,\,|\kappa(u_{t^{^{(j)}}_{i}}\!)\bm{\Sigma}_j|^{\frac{1}{2}}}$$

\subsection{Prior distribution}
\label{subsec24}
The prior distribution for the interest parameters is the following
\begin{scriptsize}

$$\pi(h,c,\bm{\theta}_1,\ldots,\bm{\theta}_l,\bm{\Sigma}_1,\ldots,\bm{\Sigma}_l,\nu)=\pi(h)\pi(c)\prod\limits_{j=1}^l \pi(\bm{\theta}_j,\bm{\Sigma}_j)\pi(\nu)=\pi(h)\pi(c)\prod\limits_{j=1}^l
\pi(\bm{\Sigma}_j)\pi(\bm{\theta}_j|\bm{\Sigma}_j)\pi(\nu),$$
\end{scriptsize}
where 
\begin{enumerate}
	\item[(1)] $\pi(c)\propto \begin{cases}1 &\text{\rm if}\quad c_1<c_2<\ldots<c_{l-1}\\ 0 &\text{\rm otherwise}\end{cases}$
	\item[(2)] $\pi(h)\propto I_{h}\{h_{\rm min},\ldots,h_{\rm max}\}$
	\item[(3)] $\pi(\bm{\Sigma}_j)$ is ${\rm W}^{-1}(\bm{\Omega}_{0j},\tau_{0j})$, in which ${\rm W}^{-1}(\bm{\Omega},\tau)$ represents the inverse Wishart
	distribution (see, for instance, \cite[section 3.4]{GN99}),
	\item[(4)] $\pi(\bm{\theta}_j|\bm{\Sigma}_j)$ is ${\rm MN}_{s_j,k}(\bm{\mu}_{0j},\bm{\Delta}_{0j},\bm{\Sigma}_j)$,\qquad and
	\item[(5)] $\pi(\nu)$ is
	\begin{itemize}
		\item for the Student-$t$ case, ${\rm Uniform}(\gamma_{0},\eta_{0})$;
		\item for the Slash case, ${\rm Gamma}(\gamma_{0},\eta_{0})$;
		\item for the contaminated normal case, $\pi(\nu_1)\pi(\nu_2)$, where $\pi(\nu_1)$ is ${\rm Beta}(\gamma_{01},\eta_{01})$ and $\pi(\nu_2)$ is 
		${\rm TGamma}(\gamma_{02},\eta_{02};(0,1))$.
		\item for the symmetric hyperbolic case, ${\rm Uniform}(\gamma_{0},\eta_{0})$.
	\end{itemize}  
\end{enumerate}
In most cases, choices are made in order to ensure conjugacy.

\subsection{MCMC estimation}
\label{subsec25}
The Gibbs sampler algorithm is as follows.
\begin{description}
	\item[Step 0:] Set the starting values of the interest parameters. 
	\begin{enumerate}
		\item[(1)] $h$ could be set at any value in $\{h_{\rm min},\ldots,h_{\rm max}\}$.
		\item[(2)] $c_j$ could be set at the quantile of order $100(j/l)\%$ of $Z_{t-h}$ for $j=1,\ldots,l-1$.
		\item[(3)] $\bm{\theta}_j$ and $\bm{\Sigma}_j$ could be set at their maximum likelihood estimates under the assumption that noise process follows a Gaussian distribution. That is,\\ $\hat{\bm{\theta}}_j=(\bm{M}_j^{\!\top}\bm{M}_j)^{\!-1}\bm{M}_j^{\!\top}\bm{Y}_j$\,\,\, and\,\,\, $\hat{\bm{\Sigma}}_j=\dfrac{1}{n_j}\!(\bm{Y}_j-\bm{M}_j\hat{\bm{\theta}}_j)^{\!\top}\!(\bm{Y}_j-\bm{M}_j\hat{\bm{\theta}}_j)$\,\,\, for\,\,\,$j=1,\ldots,l$.
		\item[(4)] The value of $\nu$ could be set so that its behavior is closest to that of the Gaussian distribution, which is
		\begin{itemize}
			\item for the Student-$t$ case, a ``large'' value, for instance, $\nu=100$;
			\item for the Slash case, a ``large'' value, for instance, $\nu=100$;
			\item for the contaminated normal case, $\nu_1=0.01$ and $\nu_2=0.99$;
			\item for the symmetric hyperbolic case, $\nu=1.85$.
		\end{itemize}
	\end{enumerate}
	\item[Step 1:] Sample $u_{t^{^{(j)}}_{i}}$ for $i=1,\ldots,n_j$ and $j=1,\ldots,l$ independently from $\pi(u_{t^{^{(j)}}_{i}}|h,c,\bm{\theta}_j,\bm{\Sigma}_j,\nu)$, which is
	\begin{itemize}
		\item for the Student-$t$ case,
		$${\rm Gamma}\!\left(\!\frac{\nu+k}{2},\frac{\nu}{2}+\frac{1}{2}\!({y}_{t^{^{(j)}}_{i}}^{\!\top}-{M}_{t^{^{(j)}}_{i}}\bm{\theta}_j)\bm{\Sigma}_j^{\!-1}({y}_{t^{^{(j)}}_{i}}^{\!\top}-{M}_{t^{^{(j)}}_{i}}\bm{\theta}_j)^{\!\top}\right);$$
		\item for the Slash case,
		$${\rm TGamma}\!\left(\!\frac{\nu+k}{2},\frac{1}{2}\!({y}_{t^{^{(j)}}_{i}}^{\!\top}-{M}_{t^{^{(j)}}_{i}}\bm{\theta}_j)\bm{\Sigma}_j^{\!-1}({y}_{t^{^{(j)}}_{i}}^{\!\top}-{M}_{t^{^{(j)}}_{i}}\bm{\theta}_j)^{\!\top};(0,1)\!\right);$$
		\item for the contaminated normal case,
		$$1 - (1-\nu_2){\rm Bernoulli}(\tau),$$
		where
		$$\tau\propto \nu_1\nu_2^{\frac{k}{2}}\exp\!\left(\!\!-\frac{\nu_2}{2}({y}_{t^{^{(j)}}_{i}}^{\!\top}-{M}_{t^{^{(j)}}_{i}}\bm{\theta}_j)\bm{\Sigma}_j^{\!-1}({y}_{t^{^{(j)}}_{i}}^{\!\top}-{M}_{t^{^{(j)}}_{i}}\bm{\theta}_j)^{\!\top}\!\right)$$
		and
		$$(1-\tau)\propto(1-\nu_1)\exp\!\left(\!\!-\frac{1}{2}({y}_{t^{^{(j)}}_{i}}^{\!\top}-{M}_{t^{^{(j)}}_{i}}\bm{\theta}_j)\bm{\Sigma}_j^{\!-1}({y}_{t^{^{(j)}}_{i}}^{\!\top}-{M}_{t^{^{(j)}}_{i}}\bm{\theta}_j)^{\!\top}\!\right);$$
		\item for the symmetric hyperbolic case,
		$${\rm GIG}\!\left(\!\frac{2-k}{2},1+({y}_{t^{^{(j)}}_{i}}^{\!\top}-{M}_{t^{^{(j)}}_{i}}\bm{\theta}_j)\bm{\Sigma}_j^{\!-1}({y}_{t^{^{(j)}}_{i}}^{\!\top}-{M}_{t^{^{(j)}}_{i}}\bm{\theta}_j)^{\!\top},\nu^2\!\right);$$
		\item for the Laplace case,
		$${\rm GIG}\!\left(\!\frac{2-k}{2},({y}_{t^{^{(j)}}_{i}}^{\!\top}-{M}_{t^{^{(j)}}_{i}}\bm{\theta}_j)\bm{\Sigma}_j^{\!-1}({y}_{t^{^{(j)}}_{i}}^{\!\top}-{M}_{t^{^{(j)}}_{i}}\bm{\theta}_j)^{\!\top},\frac{1}{4}\!\right);$$
		\item for the Gaussian case,
		$u_{t^{^{(j)}}_{i}}=1$ with probability 1.
	\end{itemize}
	\item[Step 2:] Sample $\bm{\theta}_j$ for $j=1,\ldots,l$ independently from $\pi(\bm{\theta}_j|h,c,u_j,\bm{\Sigma}_j,\nu)$, which is
	$${\rm MN}_{s_j,k}(\bm{\mu}_{j},\bm{\Delta}_j,\bm{\Sigma}^{^{(j)}}),$$
	where
	$$\bm{\mu}_j=(\bm{M}_j^{\!\top}\![\bm{\kappa}(u_j)]^{-1}\!\bm{M}_j + \bm{\Delta}_{0j}^{\!-1})^{\!-1}(\bm{M}_j^{\!\top}\![\bm{\kappa}(u_j)]^{-1}\bm{y}_j + \bm{\Delta}_{0j}^{\!-1}\bm{\mu}_{0j})$$
	and
	$$\bm{\Delta}_j=(\bm{M}_j^{\!\top}\![\bm{\kappa}(u_j)]^{-1}\!\bm{M}_j + \bm{\Delta}_{0j}^{\!-1})^{-1}.$$
	\item[Step 3:] Sample $\bm{\Sigma}_j$ for $j=1,\ldots,l$ independently from $\pi(\bm{\Sigma}_j|h,c,u_j,\bm{\theta}^{^{(j)}},\nu)$, which is an inverse Wishart distribution, denoted by
	$${\rm W}^{-1}(\bm{\Omega}_{j},\tau_j),$$
	where
	$$\bm{\Omega}_{j}=\bm{\Omega}_{0j}+(\bm{y}_j-\bm{M}_j\bm{\theta}_j)^{\!\top}\![\bm{\kappa}(u_j)]^{-1}\!(\bm{y}_j-\bm{M}_j\bm{\theta}_j)+(\bm{\theta}_j-\bm{\mu}_{0j})^{\!\top}\!\bm{\Delta}_{0j}^{-1}\!(\bm{\theta}_j-\bm{\mu}_{0j})$$
	and
	$$\tau_j=\tau_{0j}+n_j+s_j.$$
	\item[Step 4:] Sample $\nu$ from $\pi(\nu|h,c,u_1,\ldots,u_l,\bm{\theta}_1,\ldots,\bm{\theta}_l,\bm{\Sigma}_1,\ldots,\bm{\Sigma}_l)$, which is
	\begin{itemize}
		\item for the Slash case, ${\rm Gamma}(\gamma,\eta)$, where 
		$$\gamma=\gamma_{0}+\sum\limits_{j=1}^l n_j\qquad\text{and}\qquad\eta=\eta_0-\frac{1}{2}\!\sum\limits_{j=1}^l\sum\limits_{i=1}^{n_j}\log(u_{t^{^{(j)}}_{i}});$$
		\item for the contaminated normal case, $\nu_1\!\sim\!{\rm Beta}(\gamma_{1},\eta_{1})$ and $\nu_2\!\sim\!{\rm TGamma}(\gamma_{2},\eta_{2};(0,1))$, 
		where
		$$\gamma_1=\gamma_{01}+\sum\limits_{j=1}^l\sum\limits_{i=1}^{n_j}I(u_{t^{^{(j)}}_{i}}=\nu_2),\qquad\eta_1=\eta_{01}+\sum\limits_{j=1}^l\sum\limits_{i=1}^{n_j}I(u_{t^{^{(j)}}_{i}}=1),$$
		$$\gamma_2=\gamma_{02}+\frac{k}{2}\!\sum\limits_{j=1}^l\sum\limits_{i=1}^{n_j}I(u_{t^{^{(j)}}_{i}}=\nu_2)$$
		and
		$$\eta_2=\eta_{02}+\frac{1}{2}\!\sum\limits_{j=1}^l\sum\limits_{i=1}^{n_j}I(u_{t^{^{(j)}}_{i}}=\nu_2)({y}_{t^{^{(j)}}_{i}}^{\!\top}-{M}_{t^{^{(j)}}_{i}}\bm{\theta}_j)\bm{\Sigma}_j^{\!-1}({y}_{t^{^{(j)}}_{i}}^{\!\top}-{M}_{t^{^{(j)}}_{i}}\bm{\theta}_j)^{\!\top};$$
		\item for the Student-$t$ case, proportional to
		$$\tilde{\pi}(\nu)=\prod\limits_{j=1}^l\prod\limits_{i=1}^{n_j}\dfrac{\left(\!\dfrac{\nu}{2}u_{t^{^{(j)}}_{i}}\!\!\right)^{\!\!\frac{\nu}{2}}}{\Gamma\!\left(\dfrac{\nu}{2}\right)}\exp\!\!\left(\!\!-\dfrac{\nu}{2}u_{t^{^{(j)}}_{i}}\!\!\right)\!I_{\nu}(\gamma_0,\eta_0).$$
		
		Sample $\nu$ from $\pi(\nu|h,c,u_1,\ldots,u_l,\bm{\theta}_1,\ldots,\bm{\theta}_l,\bm{\Sigma}_1,\ldots,\bm{\Sigma}_l)$ is not an easy task. In order to simplify this, the distribution of 
		
		$\nu|h,c,u_1,\ldots,u_l,\bm{\theta}_1,\ldots,\bm{\theta}_l,\bm{\Sigma}_1,\ldots,\bm{\Sigma}_l$ can be discretized as described as follows. Let $m$ be a ``large'' positive integer value. The interval $(\gamma_0,\eta_0)$ is partitioned into $m$ intervals whose  limits are given by $\gamma_0=l_1<l_2<\ldots<l_{m}<l_{m+1}=\eta_0$, where $l_{r+1}=l_r + (\eta_0-\gamma_0)/m$ for $r=1,\ldots,m$. The possible values and the associated probabilities of the (discretized) random va\-ria\-ble $\nu|h,c,u_1,\ldots,u_l,\bm{\theta}_1,\ldots,\bm{\theta}_l,\bm{\Sigma}_1,\ldots,\bm{\Sigma}_l$ are given, respectively, by $(l_r + l_{r+1})/2$ and $C[\tilde{\pi}(l_r) + \tilde{\pi}(l_{r+1})](\eta_0-\gamma_0)/2m$ (area of a trapezoid) for $r=1,\ldots,m$, where $C>0$ is a constant.
		
		\item for the symmetric hyperbolic case, proportional to
		$$\tilde{\pi}(\nu)=\prod\limits_{j=1}^l\prod\limits_{i=1}^{n_j}\dfrac{\nu}{K_1(\nu)}\exp\!\!\left(\!\!-\dfrac{\nu^2}{2}u_{t^{^{(j)}}_{i}}\!\!\right)\!I_{\nu}(\gamma_0,\eta_0).$$
		
		Sample $\nu$ from $\pi(\nu|h,c,u_1,\ldots,u_l,\bm{\theta}_1,\ldots,\bm{\theta}_l,\bm{\Sigma}_1,\ldots,\bm{\Sigma}_l)$ is not an easy task. In order to simplify this, the distribution of 
		
		$\nu|h,c,u_1,\ldots,u_l,\bm{\theta}_1,\ldots,\bm{\theta}_l,\bm{\Sigma}_1,\ldots,\bm{\Sigma}_l$ can be discretized as described as follows. Let $m$ be a ``large'' positive integer value. The interval $(\gamma_0,\eta_0)$ is partitioned into $m$ intervals whose  limits are given by $\gamma_0=l_1<l_2<\ldots<l_{m}<l_{m+1}=\eta_0$, where $l_{r+1}=l_r + (\eta_0-\gamma_0)/m$ for $r=1,\ldots,m$. The possible values and the associated probabilities of the (discretized) random va\-ria\-ble $\nu|h,c,u_1,\ldots,u_l,\bm{\theta}_1,\ldots,\bm{\theta}_l,\bm{\Sigma}_1,\ldots,\bm{\Sigma}_l$ are given, respectively, by $(l_r + l_{r+1})/2$ and $C[\tilde{\pi}(l_r) + \tilde{\pi}(l_{r+1})](\eta_0-\gamma_0)/2m$ (area of a trapezoid) for $r=1,\ldots,m$, where $C>0$ is a constant.
	\end{itemize}
	
	\item[Step 5:] Sample $c=(c_1,\ldots,c_{l-1})^{\!\top}$ from $\pi(c|h,u_1,\ldots,u_l,\bm{\theta}_1,\ldots,\bm{\theta}_l,\bm{\Sigma}_1,\ldots,\bm{\Sigma}_l,\nu)$, which is proportional to

	\begin{scriptsize}

	$$\begin{cases} \prod\limits_{j=1}^l \prod\limits_{t\,:\,Z_{t-h}\in(c_{j-1},c_j]} \dfrac{\exp\!\left(\!\!-\frac{1}{2}({y}_{t^{^{(j)}}}^{\!\top}-{M}_{t^{^{(j)}}}\bm{\theta}_j)[\kappa(u_{t^{^{(j)}}}\!)\bm{\Sigma}_j]^{-1}\!({y}_{t^{^{(j)}}}^{\!\top}-{M}_{t^{^{(j)}}}\bm{\theta}_j)^{\!\top}\!\right)}{(2\pi)^{\frac{k}{2}}\,\,|\kappa(u_{t^{^{(j)}}}\!)\bm{\Sigma}_j|^{\frac{1}{2}}}&\text{\rm if}\quad c_1<c_2<\ldots<c_{l-1}\\
		0 &\text{\rm otherwise}
	\end{cases}$$
	
		\end{scriptsize}
	Since this distribution cannot be easily recognized, a Metropolis-Hastings path is required. In our proposal, a random vector drawn from a Dirichlet distribution is transformed linearly into $c$ as follows:
	$$c=\begin{bmatrix}1\\\vdots\\1\end{bmatrix}\!\!z_0 + {\bf D}\!\!\begin{bmatrix}r_1\\\vdots\\r_{l-1}\end{bmatrix}\!\!(z_1-z_0),$$
	
	where ${\bf D}$ is a square matrix of order $l-1$ whose all elements above the principal diagonal are zeros, and all other elements are ones, $r=(r_1,\ldots,r_{l})^{\!\top}\sim {\rm Dirichlet}(\kappa_1,\ldots,\kappa_l)$, $\kappa_1=(c^{\rm old}_1-z_0)/(z_1-z_0)$, $\kappa_l=(z_1-c^{\rm old}_{l-1})/(z_1-z_0)$, $\kappa_j=(c^{\rm old}_j-c^{\rm old}_{j-1})/(z_1-z_0)$ for $j=2,\ldots,l-1$, $z_0={\rm min}(Z_{t-h})$, $z_1={\rm max}(Z_{t-h})$ and $c^{\rm old}=(c^{\rm old}_1,\ldots,c^{\rm old}_{l-1})^{\!\top}$ is the value of $c$ in the previous iteration.
	
	\item[Step 6:] Sample $h$ from $\pi(h|c,u_1,\ldots,u_l,\bm{\theta}_1,\ldots,\bm{\theta}_l,\bm{\Sigma}_1,\ldots,\bm{\Sigma}_l,\nu)$, which is proportional to
	\begin{equation*}
	\resizebox{\textwidth}{!}{$\prod\limits_{j=1}^l \prod\limits_{t\,:\,Z_{t-h}\in(c_{j-1},c_j]} \dfrac{\exp\!\left(\!\!-\frac{1}{2}({y}_{t^{^{(j)}}}^{\!\top}-{M}_{t^{^{(j)}}}\bm{\theta}_j)[\kappa(u_{t^{^{(j)}}})\bm{\Sigma}_j]^{-1}({y}_{t^{^{(j)}}}^{\!\top}-{M}_{t^{^{(j)}}}\bm{\theta}_j)^{\!\top}\!\right)}{(2\pi)^{\frac{k}{2}}\,\,|\kappa(u_{t^{^{(j)}}}\!)\bm{\Sigma}_j|^{\frac{1}{2}}}I_{h}\{h_{\rm min},\ldots,h_{\rm max}\}.$}
		\end{equation*}
	
\end{description}

\subsection{Model comparison}
\label{subsec26}
\subsubsection{Deviance Information Criterion (DIC)}
\label{subsubsec261}
According to \cite{spiegelhalter02,spiegelhalter14} and \cite[section 7.2]{gelman13}, the Deviance Information Criterion (DIC) can be expressed as 
$${\rm DIC}=\widehat{\rm D} +2(\overline{\rm D}-\widehat{\rm D}),$$
where
\begin{enumerate}
	\item[$(1)$] $$\widehat{\rm D}=\sum\limits_{j=1}^l \sum\limits_{t\,:\,Z_{t-\overline{h}}\in(\overline{c}_{j-1},\overline{c}_j]} -2\log[f_{Y}({y}_{t^{^{(j)}}}|{M}_{t^{^{(j)}}}\overline{\bm{\theta}}_j,\bm{\overline{\Sigma}}_j,\overline{\nu})],$$
	in which $\overline{h}$, $\overline{c}$, $\overline{\bm{\theta}}_1,\ldots,\overline{\bm{\theta}}_l$, $\bm{\overline{\Sigma}}_1,\ldots,\bm{\overline{\Sigma}}_l$ and $\overline{\nu}$ represent, respectively, the posterior means of $h$, $c$, $\bm{\theta}_1,\ldots,\bm{\theta}_l$, $\bm{\Sigma}_1,\ldots,\bm{\Sigma}_l$ and $\nu$.
	\item[$(2)$] $$\overline{\rm D}=\dfrac{1}{G}\!\sum\limits_{g=1}^G\sum\limits_{j=1}^l \sum\limits_{t\,:\,Z_{t-h_{g}}\in(c^g_{j-1},c^g_j]} -2\log[f_{Y}({y}_{t^{^{(j)}}}|{M}_{t^{^{(j)}}}\bm{\theta}_j^g,\bm{\Sigma}_j^g,\nu_g)],$$
	in which $h_g$, $c_g$, $\bm{\theta}_1^g,\ldots,\bm{\theta}_l^g$, $\bm{\Sigma}_1^g,\ldots,\bm{\Sigma}_l^g$ and $\nu_g$ represent, respectively, the values of $h$, $c$, $\bm{\theta}_1,\ldots,\bm{\theta}_l$, $\bm{\Sigma}_1,\ldots,\bm{\Sigma}_l$ and $\nu$ in the iteration $g$ of the MCMC-type algorithm.
\end{enumerate}

\subsubsection{Watanabe-Akaike Information Criterion (WAIC)}
\label{subsubsec262}
According to \cite{w10} and \cite[section 7.2]{gelman13}, the Watanabe-Akaike Information Criterion (WAIC) can be expressed as 
$${\rm WAIC}=\widehat{\rm W} +2(\overline{\rm W}-\widehat{\rm W}),$$
where
\begin{enumerate}
	\item[$(1)$] $$\widehat{\rm W}=\sum\limits_{t}-2\log\!\left[\!\dfrac{1}{G}\!\sum\limits_{g=1}^Gf_{Y}({y}_{t^{^{(j_t^g)}}}|{M}_{t^{^{(j_t^g)}}}\bm{\theta}_{j_t^g}^g,\bm{\Sigma}_{j_t^g}^g,\nu_g)\!\right],$$
	in which $j_t^g=j\in\{1,\ldots,l\}:\,Z_{t-h_g}\in (c^g_{j-1},c^g_j]$, that is, $j_t^g$ is the regime to which $t$ belongs in iteration $g$ of the 
	MCMC-type algorithm, whereas $\bm{\theta}_{j_t^g}^g$ and $\bm{\Sigma}_{j_t^g}^g$ represent, respectively, the values of $\bm{\theta}_{j_t^g}$ and $\bm{\Sigma}_{j_t^g}$ in the iteration $g$.
	\item[$(2)$] $$\overline{\rm W}=\sum\limits_{t}\left[\!\dfrac{1}{G}\!\sum\limits_{g=1}^G-2\log[f_{Y}({y}_{t^{^{(j_t^g)}}}|{M}_{t^{^{(j_t^g)}}}\bm{\theta}_{j_t^g}^g,\bm{\Sigma}_{j_t^g}^g,\nu_g)]\!\right].$$
\end{enumerate}

\subsubsection{Normal QQ-plot}
\label{subsubsec263}
The random vectors $\bm{\epsilon}_{t^{(j)}}$ for all $t\in {\rm regime}\,j$ ($j=1,\ldots,l$) are assumed to be independent and identically distributed. Therefore, the random variables given by
\begin{align*}
	&\quad\,\,\Phi^{-1}\!\left\{\!F_{\rho}\!\!\left[\bm{\epsilon}_{t^{(j)}}^{\top}\bm{\epsilon}_{t^{(j)}}\,|\,\nu\right]\!\right\}\\
	&=\Phi^{-1}\!\left\{\!F_{\rho}\!\!\left[({Y}_{t^{^{(j)}}}^{\!\top}-{M}_{t^{^{(j)}}}\bm{\theta}_j)\bm{\Sigma}_j^{-1}({Y}_{t^{^{(j)}}}^{\!\top}-{M}_{t^{^{(j)}}}\bm{\theta}_j)^{\!\top}\,|\,\nu\right]\!\right\},
\end{align*}
become a random sample drawn from the standard normal distribution, where $F_{\rho}(\cdot|\nu)$ and $\Phi(\cdot)$ represent, respectively, the cumulative distribution functions of $\rho=\bm{\epsilon}_{t^{(j)}}^{\top}\bm{\epsilon}_{t^{(j)}}$ and standard normal. In this way, a normal QQ-plot of $r_t$ for all $t$ can be used to graphically assess the adequacy of the fitted model and to identify extreme or outlying time points. $r_t$ can be written as follows
$$r_t=\Phi^{-1}\!\left\{\!F_{\rho}\!\!\left[({y}_{t^{^{(\bar{j}_t)}}}^{\!\top}-{M}_{t^{^{(\bar{j}_t)}}}\overline{\bm{\theta}}_{\bar{j}_t})\overline{\bm{\Sigma}}_{\bar{j}_t}^{-1}\!({y}_{t^{^{(\bar{j}_t)}}}^{\!\top}-{M}_{t^{^{(\bar{j}_t)}}}\overline{\bm{\theta}}_{\bar{j}_t})^{\!\top}\,|\,\overline{\nu}\right]\!\right\},$$
where $\overline{j}_t=j\in\{1,\ldots,l\}:\,Z_{t-\overline{h}}\in (\overline{c}_{j-1},\overline{c}_j]$.

\section{Forecasting}
\label{sec3}

This section describes a $m$ step-ahead forecasting procedure based on the methodology exposed in \cite{KS13}. This methodology was used by \cite{CN17} to obtain forecasts with MTAR models when the distribution of the noise process is Gaussian. Let $t=1,\ldots,T$ be the time points corresponding to the currently observed data. Thus, the interest is to compute the forecast for $Y_t$ for the time points $t=T+1,\ldots,T+m$, where $m$ is a positive integer value. The joint predictive distribution can be written as 
\begin{scriptsize}

\begin{align*}
	f(Y_{T+1:T+m}|Y_{1:T},Z_{1:T+m},X_{1:T+m})\!&
	=\!\!\int\! f(Y_{T+1:T+m}|Y_{1:T},Z_{1:T+m},X_{1:T+m},\Theta)f(\Theta|Y_{1:T},Z_{1:T+m},X_{1:T+m})d\Theta\\
	&=\!\!\int\! f(Y_{T+1:T+m}|Y_{1:T},Z_{1:T+m},X_{1:T+m},\Theta)f(\Theta|Y_{1:T},Z_{1:T},X_{1:T})d\Theta\\
	&=\!\!\int\! \prod_{t=T+1}^{T+m}f(Y_{t}|Y_{1:t-1},Z_{1:T+m},X_{1:T+m},\Theta)f(\Theta|Y_{1:T},Z_{1:T},X_{1:T})d\Theta\\
	&=\!\!\int\! \prod_{t=T+1}^{T+m}f(Y_{t}|Y_{1:t-1},Z_{1:t},X_{1:t},\Theta)f(\Theta|Y_{1:T},Z_{1:T},X_{1:T})d\Theta,
\end{align*}
\end{scriptsize}

where $\Theta=({\rm vec}(\bm{\theta}_1)^{\!\top},{\rm vec}(\bm{\Sigma}_1)^{\!\top},\ldots,{\rm vec}(\bm{\theta}_l)^{\!\top},{\rm vec}(\bm{\Sigma}_l)^{\!\top},c^{\!\top},h,\nu)^{\!\top}$ represents the interest pa\-ra\-me\-ter vector. In the above derivation the exact values of the processes $\{X_t\}$ and $\{Z_t\}$ for $t=T+1,\ldots,T+m$ are assumed to be known. If they are not known, which is most common in practice, forecasts for $X_t$ and $Z_t$ for those time points must be obtained in advance. In addition, the above derivation is based on the following assumptions:
\begin{enumerate}
	\item[$(1)$] $f(\Theta|Y_{1:T},Z_{1:T+m},X_{1:T+m})$ coincide with $f(\Theta|Y_{1:T},Z_{1:T},X_{1:T})$, which represents the posterior distribution of $\Theta$.
	\item[$(2)$] $f(Y_{t}|Y_{1:t-1},Z_{1:T+m},X_{1:T+m})$ coincide with $f(Y_{t}|Y_{1:t-1},Z_{1:t},X_{1:t})$ for all $t\in\{T+1,\ldots,T+m\}$. Note that $f(Y_{t}|Y_{1:t-1},Z_{1:t},X_{1:t})$ corresponds to a distribution of the class of multivariate Gaussian variance mixtures as the equation of the MTAR model can be written as a multivariate multiple regression in the same way as in (\ref{LMMTAR}), that is,
	
	$${Y}_{t}=(M_{t^{^{(j)}}}\bm{\theta}_j)^{\!\top}+\bm{\Sigma}_j^{\frac{1}{2}}\epsilon_t,$$
	where $j$ is such that $Z_{t-h}\in(c_{j-1},c_j]$ and the distribution of $\epsilon_t$ belongs to the class of multivariate normal variance mixtures.
\end{enumerate}

\noindent The following is a description of the forecasting procedure.

\begin{description}
	\item[Step 0:] For $Z_t$ and $X_t$, forecast the time points $t=T+1,\ldots,T+m$. In addition, values of the structural parameters $l$, $p$, $q$ and $d$ must be set.\\
	
	\item[Steo 1:] For $g$ from 1 to $G$, generate $\bm{\Theta}^{g}$ according to the MCMC-type algorithm described above.
	\item[Step 2:] For $g$ from 1 to $G$, generate $y_{t}^g$ according to $f_{Y_t}({y}_{\!t^{^{(j_t^g)}}}|{M}_{\!t^{^{(j_t^g)}}}\bm{\theta}_{j_t^g}^g,\bm{\Sigma}_{j_t^g}^g,\nu_g)$, where $t=T+1$. The forecast for $Y_{T+1}$ can then be computed as
	$$\hat{\rm E}(Y_{T+1}|Y_{1:T},Z_{1:T+1},X_{1:T+1})=\dfrac{1}{G}\!\sum\limits_{g=1}^G y_{T+1}^g$$
	\item[Step 3:] For $g$ from 1 to $G$, use $y_{_{T+1}}^g$ to generate $y_{t}^g$ according to $f_{Y_t}({y}_{\!t^{^{(j_t^g)}}}|{M}_{\!t^{^{(j_t^g)}}}\bm{\theta}_{j_t^g}^g,\bm{\Sigma}_{j_t^g}^g,\nu_g)$, where $t=T+2$. The forecast for $Y_{T+2}$ can then be computed as
	$$\hat{\rm E}(Y_{T+2}|Y_{1:T},Z_{1:T+2},X_{1:T+2})=\dfrac{1}{G}\!\sum\limits_{g=1}^G y_{T+2}^g.$$
	\item[Step 4:] For $g$ from 1 to $G$, use $y_{_{T+1}}^g$ and $y_{_{T+2}}^g$ to generate $y_{t}^g$ according to $f_{Y_t}({y}_{\!t^{^{(j_t^g)}}}|{M}_{\!t^{^{(j_t^g)}}}\bm{\theta}_{j_t^g}^g,\bm{\Sigma}_{j_t^g}^g,\nu_g)$, where $t=T+3$. The forecast for $Y_{T+3}$ can then be computed as
	$$\hat{\rm E}(Y_{T+3}|Y_{1:T},Z_{1:T+3},X_{1:T+3})=\dfrac{1}{G}\!\sum\limits_{g=1}^G y_{T+3}^g.$$
	$\vdots$\\
	\item[Step $\bm{m}$:] For $g$ from 1 to $G$, use $y_{_{T+1}}^g,y_{_{T+2}}^g,\ldots,y_{_{T+m-1}}^g$ to generate $y_{t}^g$ according to $f_{Y_t}({y}_{\!t^{^{(j_t^g)}}}|{M}_{\!t^{^{(j_t^g)}}}\bm{\theta}_{j_t^g}^g,\bm{\Sigma}_{j_t^g}^g,\nu_g)$, where $t=T+m$. The forecast for $Y_{T+m}$ 
	can then be computed as
	$$\hat{\rm E}(Y_{T+m}|Y_{1:T},Z_{1:T+m},X_{1:T+m})=\dfrac{1}{G}\!\sum\limits_{g=1}^G y_{T+m}^g.$$
\end{description}
Thus, $y_{T+1}^g,\ldots,y_{T+m}^g$ for $g=1,\ldots,G$ is a sample of size $G$ from the joint predictive distribution, which can also be used to calculate credibility intervals for forecasts. 
\newpage

\section{Simulation Experiment}
\label{sec4}
Simulation experiments were conducted to evaluate the estimation and forecasting proposal. Simulations of three-dimensional MTAR models with two regimes and two-dimensional MTAR models with three regimes were carried out using sample sizes of 300 and 1000 and all distributions described above for the noise process. From that, the proportion of times the $95\%$ credible and the $95\%$ prediction intervals capture the true parameter values and the output values of step-ahead, respectively, are computed based on $1000$ replications. The above is based on chains of length $1500$ with a burn-in period of $500$ for all distributions except the symmetric hyperbolic one, for which chains of length $2000$ were used. The library {\tt mtarm} (\cite{VCR24}) of the language and environment for statistical computing {\tt R} (\cite{R2024}) was used for implementation of estimation and forecasting in MTAR models. That library can be found at \url{https://cran.r-project.org/web/packages/mtarm/index.html}. It is important to point out that the hyperparameters of prior distributions were set to get non-informative priors distributions. 

We call $M_1$ the three-dimension MTAR model with 2 regimes, which is defined as follows:
\begin{scriptsize}
\begin{equation}
	Y_t=
	\begin{cases}
		\begin{bmatrix}
			1\\
			-2\\
			6
		\end{bmatrix} 
		+
		\begin{bmatrix}
			0.1&  0.6&  0.4\\
			-0.4&  0.5& -0.7\\
			0.2&  0.6& -0.3
		\end{bmatrix} 
		Y_{t-1}
		+
		\begin{bmatrix}
			0.6& -0.5\\
			-0.4&  0.6\\
			0.1&  0.3
		\end{bmatrix} 
		X_{t-1}
		+
		\begin{bmatrix}
			1&    0&    0\\
			0&    1&    0\\
			0&    0&    1\\
		\end{bmatrix}^{\!\frac{1}{2}}\!
		\epsilon_t\quad & \text{when}\quad Z_{t}\leq 0, \\
		
		\begin{bmatrix}
			0\\
			0\\
			0
		\end{bmatrix} 
		+
		\begin{bmatrix}
			0.3&  0.5& -0.5\\
			0.2&  0.7& -0.1\\
			0.3& -0.4&  0.6
		\end{bmatrix} 
		Y_{t-1}
		+
		\begin{bmatrix}
			0.3&  0.0&  0.0\\
			0.0& -0.6&  0.0\\
			0.0&  0.0&  0.5\\
		\end{bmatrix} 
		Y_{t-2}
		+
		\begin{bmatrix}
			1.5&    0&    0\\
			0.0&    1&    0\\
			0.0&    0&    2
		\end{bmatrix}^{\!\frac{1}{2}}\!
		\epsilon_t\quad & \text{when}\quad  Z_{t}> 0.\\
	\end{cases}
	\label{M1}
\end{equation}
\end{scriptsize}
We set that covariate and  threshold processes  $\{(X_t^{\!\top},Z_t)^{\!\top}\}$ are jointly generated  for a stationary three-dimensional ${\rm VAR}(1)$ defined as follows
\begin{equation}
	\begin{bmatrix}
		X_{1t}\\
		X_{2t}\\
		Z_t
	\end{bmatrix} =\begin{bmatrix}
		0.24&  0.48& -0.12\\
		0.46& -0.36&  0.10\\
		-0.12& -0.47&  0.58
	\end{bmatrix}\begin{bmatrix}
		X_{1t-1}\\
		X_{2t-1}\\
		Z_{t-1}
	\end{bmatrix}+a_t,
	\label{thresholdprocessM1}
\end{equation}
where the error process $\{a_t\}$ is a three-dimensional independent and identically distributed Gaussian process whose variance-covariance matrix is given by  $2\,{\bf I}_3$. The threshold value is set to $c_1=0$ corresponding to the $50\%$ quantile of $Z_t$. The error process $\{ \epsilon_t\}$ corresponds to each of the normal variance mixture distributions listed in Table \ref{smgd} plus the Gaussian one.

Similarly, we call $M_2$ the two-dimension MTAR model with 3 regimes, which is defined as follows:
\begin{equation}
	Y_t=
	\begin{cases}
		\begin{bmatrix}
			2\\
			1
		\end{bmatrix} 
		+
		\begin{bmatrix}
			0.8 & 0.0\\
			-0.2 & 0.5
		\end{bmatrix} 
		Y_{t-1}
		+
		\begin{bmatrix}
			1  &  0\\
			0  &  4
		\end{bmatrix}^{\!\frac{1}{2}}\! 
		\epsilon_t\quad & \text{when}\quad Z_{t-1}\leq 1.95
		, \\
		
		\begin{bmatrix}
			0.4\\
			-0.2
		\end{bmatrix} 
		+
		\begin{bmatrix}
			0.3 & 0.0 \\
			0.0 &-0.6
		\end{bmatrix} 
		Y_{t-1}
		+
		\begin{bmatrix}
			1  &  0\\
			0  &  1
		\end{bmatrix}^{\!\frac{1}{2}}\! 
		\epsilon_t\quad & \text{when}\quad 1.95
		< Z_{t-1} \leq 3.02
		,\\
		
		\begin{bmatrix}
			-3\\
			0
		\end{bmatrix} 
		+
		\begin{bmatrix}
			0.6 & 0.0 \\
			-0.2 &0.8
		\end{bmatrix} 
		Y_{t-1}
		+
		\begin{bmatrix}
			2  &  0\\
			0  &  1
		\end{bmatrix}^{\!\frac{1}{2}}\! 
		\epsilon_t\quad & \text{when}\quad  Z_{t-1}> 
		3.02
		
	\end{cases}
	\label{M2}
\end{equation}
We set that the threshold process $\{Z_t\}$ is a stationary ${\rm AR}(1)$ defined as follows
\begin{equation}
	Z_t=1+0.6\,Z_{t-1}+a_t,
	\label{thresholdprocessM"}
\end{equation}
where the error process $\{a_t\}$ is an independent and identically distributed standard Gaussian process. The threshold vector is 
$c=(1.95,3.02)^{\!\top}$, which corresponds to the theoretical $33\%$ and $66\%$ quantiles of an univariate Gaussian distribution with mean $2.5$ and variance $1.5625$. The error process $\{ \epsilon_t\}$ corresponds to each of the normal variance mixture distributions listed in Table \ref{smgd} plus the Gaussian one.

\subsection{Estimation}
\label{subsec41}
The percentage of times the true parameter values are contained in their respective $95\%$ credible intervals when the sample size is 1000 and the models $M_1$ and $M_2$ are considered are presented in Tables \ref{CIM11000} and \ref{CIM21000}, respectively. That percentage is usually close to the theoretical value, except for the thresholds ($c$) and the extra parameter ($\nu$). However, the relative or absolute bias for those parameters (tables \ref{BiasM11000} and \ref{BiasM21000}) is usually ``small'' except in the extra parameter for the contaminated normal distribution, suggesting that the estimation procedure performed well. Similar results are obtained when the sample size is 300 (see Appendix \ref{appendix:A}).

\begin{table}[!ht]
	\centering
	\caption{Percentage of times that the true parameter values lie on the $95\%$ credible intervals considering $M_1$ with sample size $1000$. For the delay parameter $h$ is considered the percentage of times that posterior mode coincides with the true delay value.}\label{CIM11000}
	\resizebox{\columnwidth}{!}{
		\begin{tabular}{c|c|c|c|c|c|c}\hline       
			&  \multicolumn{6}{c}{Distribution of the noise process} \\\cline{2-7}
			& \multirow{2}{*}{Gaussian}&  \multirow{2}{*}{Student-$t$($\nu\!=\!3$)} & \multirow{2}{*}{Slash($\nu\!=\!6$)} & Contaminated & Symmetric & \multirow{2}{*}{Laplace} \\ 
			& & &  & normal      & hyperbolic& \\
			& & &  & ($\nu_1\!=\!0.05,\nu_2\!=\!0.1$)& ($\nu\!=\!0.11$)          & \\\hline
			Regime 1 & & & & & & \\
			$\phi_0^{(1)}$ & $\begin{bmatrix}
				93.4\\
				95.7\\
				94.4
			\end{bmatrix}$ &$\begin{bmatrix}
				94.0\\
				94.7\\
				93.1\\
			\end{bmatrix}$ &$\begin{bmatrix}
				95.3\\
				95.5\\
				93.7
			\end{bmatrix}$ & $\begin{bmatrix}
				95.2\\
				94.3\\
				93.9
			\end{bmatrix}$ & $\begin{bmatrix}
				93.6\\
				94.6\\
				94.9
			\end{bmatrix}$ & $\begin{bmatrix}
				94.5\\
				94.1\\
				93.5
			\end{bmatrix}$\\ \hline
			
			$\bm{\phi}_1^{(1)}$      & $\begin{bmatrix}
				94.5& 94.7& 95.0\\
				94.3& 94.3& 94.8\\
				94.6& 94.1& 94.6
			\end{bmatrix}$ & $\begin{bmatrix}
				93.7& 93.8& 95.8\\
				94.7& 95.2& 92.9\\
				95.3& 93.7& 94.4
			\end{bmatrix}$ & $\begin{bmatrix}
				93.8& 94.9& 94.0\\
				94.2& 94.2& 94.5\\
				95.0& 93.9& 93.3
			\end{bmatrix}$& $\begin{bmatrix}
				93.8& 95.3& 95.5\\
				95.0& 94.8& 94.4\\
				94.0& 95.0& 94.2
			\end{bmatrix}$ & $\begin{bmatrix}
				94.3& 94.2& 94.4\\
				94.5& 95.2& 94.1\\
				93.1& 94.8& 93.6
			\end{bmatrix}$ & $\begin{bmatrix}
				94.8& 93.7& 93.9\\
				93.8& 94.4& 94.3\\
				93.4& 93.1& 94.2
			\end{bmatrix}$\\ \hline
			$\bm{\beta}_1^{(1)}$      & $\begin{bmatrix}
				95.0& 94.5\\
				94.6& 94.2\\
				94.1& 94.2
			\end{bmatrix}$ & $\begin{bmatrix}
				94.6& 94.4\\
				93.6& 94.7\\
				95.6& 93.8
			\end{bmatrix}$ & $\begin{bmatrix}
				94.2& 95.0\\
				93.8& 95.7\\
				94.3& 94.4
			\end{bmatrix}$& $\begin{bmatrix}
				94.3& 94.9\\
				94.2& 94.0\\
				94.1& 93.6
			\end{bmatrix}$ & $\begin{bmatrix}
				95.0& 94.8\\
				93.8& 94.7\\
				93.5& 95.2
			\end{bmatrix}$ & $\begin{bmatrix}
				95.2& 93.6\\
				93.4& 94.5\\
				94.2& 94.7
			\end{bmatrix}$
			\\\hline
			$\bm{\Sigma}_1$  & $\begin{bmatrix}
				93.1& 93.6& 94.2\\
				93.6& 94.3& 95.0\\
				94.2& 95.0& 93.3
			\end{bmatrix}$ & $\begin{bmatrix}
				87.5& 94.9& 94.2\\
				94.9& 87.5& 95.0\\
				94.2& 95.0& 88.0
			\end{bmatrix}$ &$\begin{bmatrix}
				82.8& 94.8& 95.6\\
				94.8& 81.1& 95.2\\
				95.6& 95.2& 83.9
			\end{bmatrix}$ & $\begin{bmatrix}
				92.9& 94.8& 95.0\\
				94.8& 91.7& 94.2\\
				95.0& 94.2& 93.4
			\end{bmatrix}$ & $\begin{bmatrix}
				95.5& 95.2& 96.1\\
				95.2& 95.7& 95.6\\
				96.1& 95.6& 96.0
			\end{bmatrix}$ & $\begin{bmatrix}
				90.3& 93.6& 94.4\\
				93.6& 90.9& 95.1\\
				94.4& 95.1& 90.6
			\end{bmatrix}$\\ \hline
			Regime 2 & & & & & & \\
			$\phi_0^{(2)}$ & $\begin{bmatrix}
				93.5\\
				94.5\\
				94.9\\
			\end{bmatrix}$ & $\begin{bmatrix}
				93.9\\
				94.0\\
				94.4
			\end{bmatrix}$ & $\begin{bmatrix}
				93.3\\
				94.0\\
				94.9
			\end{bmatrix}$ & $\begin{bmatrix}
				94.2\\
				93.6\\
				93.5
			\end{bmatrix}$ & $\begin{bmatrix}
				92.8\\
				95.4\\
				93.3
			\end{bmatrix}$ & $\begin{bmatrix}
				94.9\\
				93.9\\
				94.5
			\end{bmatrix}$\\ \hline
			
			$\bm{\phi}_1^{(2)}$      & $\begin{bmatrix}
				94.0& 94.9& 94.7\\
				95.0& 95.5& 95.0\\
				94.5& 93.7& 93.4
			\end{bmatrix}$ & $\begin{bmatrix}
				93.7& 93.1& 93.9\\
				93.8& 94.6& 94.4\\
				94.0& 94.8& 94.4
			\end{bmatrix}$ & $\begin{bmatrix}
				93.7& 94.0& 92.4\\
				93.8& 94.4& 94.5\\
				93.5& 95.9& 94.6
			\end{bmatrix}$ & $\begin{bmatrix}
				94.6& 94.6& 94.7\\
				95.4& 94.5& 94.9\\
				95.9& 93.5& 94.1
			\end{bmatrix}$ & $\begin{bmatrix}
				93.3& 94.4& 95.2\\
				93.7& 94.1& 94.5\\
				93.5& 95.6& 94.0
			\end{bmatrix}$ & $\begin{bmatrix}
				94.6& 94.0& 95.0\\
				94.5& 94.7& 93.8\\
				94.2& 92.7& 93.9
			\end{bmatrix}$\\ \hline
			
			$\bm{\phi}_2^{(2)}$      & $\begin{bmatrix}
				94.2& 93.4& 95.8\\
				95.0& 95.3& 94.8\\
				94.4& 94.6& 94.1
			\end{bmatrix}$ & $\begin{bmatrix}
				94.5& 93.4& 93.3\\
				93.5& 94.5& 94.4\\
				94.7& 95.2& 94.9
			\end{bmatrix}$ & $\begin{bmatrix}
				94.7& 94.6& 94.5\\
				94.9& 94.1& 94.5\\
				94.3& 94.8& 95.6
			\end{bmatrix}$ & $\begin{bmatrix}
				93.6& 94.1& 94.0\\
				93.2& 93.6& 93.7\\
				94.1& 93.5& 93.7
			\end{bmatrix}$ & $\begin{bmatrix}
				94.1& 94.1& 94.6\\
				94.0& 94.6& 95.3\\
				95.6& 94.4& 95.3
			\end{bmatrix}$ & $\begin{bmatrix}
				95.6& 94.3& 94.9\\
				94.2& 94.4& 94.5\\
				94.9& 94.3& 93.3
			\end{bmatrix}$\\ \hline

			$\bm{\Sigma}_2$  & $\begin{bmatrix}
				92.6& 93.1& 93.8\\
				93.1& 92.5& 94.9\\
				93.8& 94.9& 94.1
			\end{bmatrix}$ & $\begin{bmatrix}
				85.9& 94.2& 93.1\\
				94.2& 86.2& 94.5\\
				93.1& 94.5& 86.1
			\end{bmatrix}$ & $\begin{bmatrix}
				80.2& 95.4& 95.9\\
				95.4& 84.7& 94.8\\
				95.9& 94.8& 81.0
			\end{bmatrix}$ & $\begin{bmatrix}
				92.7& 94.2& 93.8\\
				94.2& 92.2& 95.6\\
				93.8& 95.6& 93.8
			\end{bmatrix}$ & $\begin{bmatrix}
				95.1& 95.8& 95.4\\
				95.8& 94.5& 95.7\\
				95.4& 95.7& 94.4
			\end{bmatrix}$ & $\begin{bmatrix}
				89.4& 94.4& 94.9\\
				94.4& 89.4& 94.5\\
				94.9& 94.5& 90.1
			\end{bmatrix}$\\ \hline
			$c$ & 42.2 & 36.6 &30.9 & 35.3 & 42.4 & 40.4
			\\
			$h$ & 100  & 100 & 100 & 100 &100 & 100 \\
			$\nu$ &&78.2 & 67.9& $\begin{bmatrix}
				93.3\\ 91.2
			\end{bmatrix}$ & 1 &  \\\hline
		\end{tabular}
	}
\end{table}

\begin{table}[!ht]
	\centering
	\caption{Bias for the threshold vector ($c$) and the relative bias$\times100$ for extra parameter ($\nu$) considering $M_1$ with sample size $1000$.}
	\label{BiasM11000}
	\resizebox{\columnwidth}{!}{
		\begin{tabular}{c|c|c|c|c|c|c}\hline       
			&  \multicolumn{6}{c}{Distribution of the noise process} \\\cline{2-7}
			& \multirow{2}{*}{Gaussian}&  \multirow{2}{*}{Student-$t$($3$)} & \multirow{2}{*}{Slash($6$)} & Contaminated & Symmetric & \multirow{2}{*}{Laplace} \\ 
			& & &  & normal      & hyperbolic& \\
			& & &  & ($\nu_1\!=\!0.05,\nu_2\!=\!0.1$)& ($\nu\!=\!0.11$)          & \\\hline
			$c$ & -0.00144384 & -0.00004438 & -0.000008.71 & -0.00041812 & -0.00048561 & -0.00051967\\
			$\nu$ & & 4.236  & 6.358    & $\begin{bmatrix}
				70.816\\ 85.110
			\end{bmatrix}$ & 0.621  & \\
	\end{tabular}}
\end{table}

\begin{table}[!ht]
	\centering
	\caption{Percentage of times that the true parameter values lie on the $95\%$ credible intervals considering $M_2$ with sample size $1000$. For the delay parameter $h$ is considered the percentage of times that posterior mode coincides with the true delay value.}\label{CIM21000}
\resizebox{\columnwidth}{!}{
		\begin{tabular}{c|c|c|c|c|c|c}\hline       
			&  \multicolumn{6}{c}{Distribution of the noise process} \\\cline{2-7}
			& \multirow{2}{*}{Gaussian}&  \multirow{2}{*}{Student-$t$($\nu\!=\!5$)} & \multirow{2}{*}{Slash($\nu\!=\!4$)} & Contaminated & Symmetric & \multirow{2}{*}{Laplace} \\ 
			& & &  & normal& hyperbolic& \\
			& & &  & ($\nu_1\!=\!0.08,\nu_2\!=\!0.012$)& ($\nu\!=\!0.12$)& \\\hline
			Regime 1 & & & & & & \\
			$\phi_0^{(1)}$ & $\begin{bmatrix}
				90.9\\
				83.8
			\end{bmatrix}$ &$\begin{bmatrix}
				92.8\\
				91.2
			\end{bmatrix}$ &$\begin{bmatrix}
				92.1\\
				88.8
			\end{bmatrix}$ & $\begin{bmatrix}
				91.3\\
				91.2
			\end{bmatrix}$ & $\begin{bmatrix}
				93.9\\
				94.4
			\end{bmatrix}$ & $\begin{bmatrix}
				92.9 \\
				93.4
			\end{bmatrix}$\\ \hline
			
			$\bm{\phi}_1^{(1)}$      &$\begin{bmatrix}
				90.0 &95.0\\
				90.5 &85.8
			\end{bmatrix}$ & $\begin{bmatrix}
				92.8& 94.3\\
				94.4& 91.6\\
			\end{bmatrix}$ & $\begin{bmatrix}
				93.6 & 95.0\\
				93.4 & 89.9
			\end{bmatrix}$& $\begin{bmatrix}
				95.4& 93.3\\
				93.8& 93.0
			\end{bmatrix}$ & $\begin{bmatrix}
				92.9 &94.1 \\
				96.3 &91.7
			\end{bmatrix}$ & $\begin{bmatrix}
				93.7 & 94.7\\
				94.7 & 91.1
			\end{bmatrix}$\\ \hline
			$\bm{\Sigma}_1$  & $\begin{bmatrix}
				71.6 &89.9\\
				89.9 &84.6
			\end{bmatrix}$ & $\begin{bmatrix}
				85.9& 94.2\\
				94.2& 86.0\\
			\end{bmatrix}$ &$\begin{bmatrix}
				74.7 &95.4\\
				95.4 &70.5
			\end{bmatrix}$ & $\begin{bmatrix}
				89.2& 92.2\\
				92.2& 89.7
			\end{bmatrix}$ & $\begin{bmatrix}
				96.1 & 95.3\\
				95.3 & 95.8
			\end{bmatrix}$ & $\begin{bmatrix}
				92.5 &95.0\\
				95.0 &92.5
			\end{bmatrix}$\\ \hline
			Regime 2 & & & & & & \\
			$\phi_0^{(2)}$ & $\begin{bmatrix}
				78.3\\
				78.1
			\end{bmatrix}$ & $\begin{bmatrix}
				89.9\\
				93.3
			\end{bmatrix}$ & $\begin{bmatrix}
				87.1\\
				93.3
			\end{bmatrix}$ & $\begin{bmatrix}
				88.8\\
				92.6
			\end{bmatrix}$ & $\begin{bmatrix}
				94.8\\
				95.5
			\end{bmatrix}$ & $\begin{bmatrix}
				93.9\\
				93.6
			\end{bmatrix}$\\ \hline
			
			$\bm{\phi}_1^{(2)}$      & $\begin{bmatrix}
				80.0 &95.2\\
				90.6 &73.5
			\end{bmatrix}$ & $\begin{bmatrix}
				93.7& 94.2\\
				94.5& 91.4\\
			\end{bmatrix}$ & $\begin{bmatrix}
				92.8 &95.1 \\
				96.7 &92.3
			\end{bmatrix}$ & $\begin{bmatrix}
				94.0& 95.3\\
				94.3& 92.4
			\end{bmatrix}$ & $\begin{bmatrix}
				92.4 &95.2\\
				95.3 &90.3
			\end{bmatrix}$ & $\begin{bmatrix}
				94.8 & 95.0\\
				95.7 & 90.5
			\end{bmatrix}$\\ \hline
			$\bm{\Sigma}_2$  & $\begin{bmatrix}
				49.0 &73.0\\
				73.0 &47.9
			\end{bmatrix}$ & $\begin{bmatrix}
				89.4& 91.9\\
				91.9& 89.7\\
			\end{bmatrix}$ & $\begin{bmatrix}
				88.3 & 92.2\\
				92.2 & 88.9
			\end{bmatrix}$ & $\begin{bmatrix}
				87.7& 88.2\\
				88.2& 80.8
			\end{bmatrix}$ & $\begin{bmatrix}
				96.9 &96.2\\
				96.2 &96.3
			\end{bmatrix}$ & $\begin{bmatrix}
				87.1 &92.8\\
				92.8 &66.9
			\end{bmatrix}$\\ \hline
			Regime 3 & & & & & & \\
			$\phi_0^{(3)}$ & $\begin{bmatrix}
				88.4\\
				91.7
			\end{bmatrix}$ & $\begin{bmatrix}
				92.0\\
				93.6
			\end{bmatrix}$ &$\begin{bmatrix}
				92.3\\
				93.8
			\end{bmatrix}$ & $\begin{bmatrix}
				92.5\\
				94.2
			\end{bmatrix}$ & $\begin{bmatrix}
				94.4\\
				95.8
			\end{bmatrix}$ & $\begin{bmatrix}
				92.9\\
				93.6
			\end{bmatrix}$\\ \hline
			
			$\bm{\phi}_1^{(3)}$      & $\begin{bmatrix}
				95.0 &95.4\\
				94.9 &86.5
			\end{bmatrix}$ & $\begin{bmatrix}
				96.0& 95.0\\
				93.7& 92.5
			\end{bmatrix}$ & $\begin{bmatrix}
				94.7& 94.4\\
				94.3&  92.3
			\end{bmatrix}$ & $\begin{bmatrix}
				94.7& 93.3\\
				93.4& 94.1
			\end{bmatrix}$ & $\begin{bmatrix}
				95.2 &93.9\\
				93.6 &94.0
			\end{bmatrix}$ & $\begin{bmatrix}
				94.7 & 95.2\\
				94.2 & 92.4
			\end{bmatrix}$\\ \hline
			$\bm{\Sigma}_3$  & $\begin{bmatrix}
				83.3 &83.3\\
				83.3 &76.3
			\end{bmatrix}$ & $\begin{bmatrix}
				84.7& 93.5\\
				93.5& 82.2\\
			\end{bmatrix}$ &$\begin{bmatrix}
				71.4 &94.9\\
				94.9 &69.8
			\end{bmatrix}$ & $\begin{bmatrix}
				88.7& 93.5\\
				93.5& 85.3
			\end{bmatrix}$ & $\begin{bmatrix}
				95.9 &96.2\\
				96.2 &95.7
			\end{bmatrix}$ & $\begin{bmatrix}
				94.0 &94.8\\
				94.8 &85.9
			\end{bmatrix}$\\ \hline
			$c$ & $\begin{bmatrix}
				19.2\\
				18.8
			\end{bmatrix}$ & $\begin{bmatrix}
				6.0\\
				5.7
			\end{bmatrix}$ &$\begin{bmatrix}
				7.9\\
				4.8
			\end{bmatrix}$ & $\begin{bmatrix}
				6.6\\
				5.9
			\end{bmatrix}$ & $\begin{bmatrix}
				6.1\\
				6.5
			\end{bmatrix}$& $\begin{bmatrix}
				6.5\\
				4.3
			\end{bmatrix}$ 
			\\
			$h$ &100& 100& 100 & 100 & 100 & 100\\
			$\nu$ & &39.1 &29.1 & $\begin{bmatrix}
				77.4\\
				51.0
			\end{bmatrix}$ & 97.0 &  \\\hline
	\end{tabular} }
\end{table}

\begin{table}[!ht]
	\centering
	\caption{Relative bias for the threshold vector ($c$) and the extra parameter ($\nu$) considering $M_2$ with sample size $1000$.}
	\label{BiasM21000}
	\resizebox{\columnwidth}{!}{
		\begin{tabular}{c|c|c|c|c|c|c}\hline       
			&  \multicolumn{6}{c}{Distribution of the noise process} \\\cline{2-7}
			& \multirow{2}{*}{Gaussian}&  \multirow{2}{*}{Student-$t$($\nu\!=\!5$)} & \multirow{2}{*}{Slash($\nu\!=\!4$)} & Contaminated & Symmetric & \multirow{2}{*}{Laplace} \\ 
			& & &  & normal& hyperbolic& \\
			& & &  & ($\nu_1\!=\!0.08,\nu_2\!=\!0.012$)& ($\nu\!=\!0.12$)& \\\hline
			$c$ & $\begin{bmatrix}
				0.337\\
				0.227
			\end{bmatrix}$ & $\begin{bmatrix}
				0.381\\
				0.293
			\end{bmatrix}$ & $\begin{bmatrix}
				0.479\\
				0.432
			\end{bmatrix}$ & $\begin{bmatrix}
				0.388\\
				0.425
			\end{bmatrix}$ & $\begin{bmatrix}
				0.295\\
				0.331
			\end{bmatrix}$& $\begin{bmatrix}
				0.560\\
				0.479
			\end{bmatrix}$\\
			$\nu$ & & 19.991 & 20.729 & $\begin{bmatrix}
				65.622\\
				1334.021
			\end{bmatrix}$ & 4.809 & \\
	\end{tabular}}
\end{table}

Furthermore, simulation experiments indicate that the estimation procedure performs better with fewer regimes since the number of observations used for estimating parameters in each regime decreases as the number of regimes increases for a fixed sample size.

\newpage
\subsection{Forecasting}
\label{subsec42}
In this case, we present in tables(\ref{FORECIM11000}, \ref{FORECIM21000}) the percentage of times that the true values of the vector $\bm{y}_{t+h}$ for $h=1,\cdots,10$ lies on individual credible intervals. We can observe that percentages of the individual credible intervals are kept close to the theoretical individual credibility, which is $95\%$ for all distributions and the models $M_1$ and $M_2$. 



\begin{table}[!ht]
	\centering
	\caption{Percentage of times that the true values lie on $95\%$ credible intervals considering $M_1$ with sample size $1000$.}\label{FORECIM11000}
\resizebox{\columnwidth}{!}{
		\begin{tabular}{c|c|c|c|c|c|c}\hline       
			&  \multicolumn{6}{c}{Distribution of the noise process} \\\cline{2-7}
			step ahead & \multirow{2}{*}{Gaussian}&  \multirow{2}{*}{Student-$t$($\nu\!=\!3$)} & \multirow{2}{*}{Slash($\nu\!=\!6$)} & Contaminated & Symmetric & \multirow{2}{*}{Laplace} \\ 
			& & &  & normal      & hyperbolic& \\
			& & &  & ($\nu_1\!=\!0.05,\nu_2\!=\!0.1$)& ($\nu\!=\!0.11$)          & \\\hline
			$y_{T+1}$ & $\begin{bmatrix}
				94.7\\ 95.1\\ 94.1
			\end{bmatrix}$ &$\begin{bmatrix}
				95.2\\94.8\\94.6
			\end{bmatrix}$ &$\begin{bmatrix}
				93.3\\ 94.2\\ 94.2
			\end{bmatrix}$ & $\begin{bmatrix}
				95.2\\ 95.5\\ 94.7
			\end{bmatrix}$ & $\begin{bmatrix}
				94.0\\ 94.2\\ 93.5
			\end{bmatrix}$ & $\begin{bmatrix}
				94.2\\ 92.5\\ 94.7
			\end{bmatrix}$\\ \hline
			$y_{T+2}$ & $\begin{bmatrix}
				94.5\\ 94.5\\ 94.0
			\end{bmatrix}$ &$\begin{bmatrix}
				95.0\\ 93.5\\ 94.2
			\end{bmatrix}$ &$\begin{bmatrix}
				94.3\\ 94.6\\ 93.6
			\end{bmatrix}$ & $\begin{bmatrix}
				94.0\\ 94.0\\ 94.6
			\end{bmatrix}$ & $\begin{bmatrix}
				93.7\\ 94.4\\ 94.5
			\end{bmatrix}$ & $\begin{bmatrix}
				93.6\\ 94.1\\ 94.0
			\end{bmatrix}$\\ \hline
			$y_{T+3}$ & $\begin{bmatrix}
				95.5\\ 95.0\\ 92.8
			\end{bmatrix}$ &$\begin{bmatrix}
				93.9\\ 93.5\\ 93.3
			\end{bmatrix}$ &$\begin{bmatrix}
				95.5\\ 95.6\\ 94.3
			\end{bmatrix}$ & $\begin{bmatrix}
				95.6\\ 94.4\\ 94.8
			\end{bmatrix}$ & $\begin{bmatrix}
				94.3\\ 94.0\\ 93.9
			\end{bmatrix}$ & $\begin{bmatrix}
				94.5\\ 93.6\\ 93.9
			\end{bmatrix}$\\ \hline
			$y_{T+4}$ & $\begin{bmatrix}
				93.8\\ 94.6\\ 93.1
			\end{bmatrix}$ &$\begin{bmatrix}
				94.2\\ 93.8\\ 95.0
			\end{bmatrix}$ &$\begin{bmatrix}
				95.2\\ 95.2\\ 94.7
			\end{bmatrix}$ & $\begin{bmatrix}
				94.8\\ 94.0\\ 93.6
			\end{bmatrix}$ & $\begin{bmatrix}
				93.3\\ 93.7\\ 94.6
			\end{bmatrix}$ & $\begin{bmatrix}
				94.8\\ 93.3\\ 94.2
			\end{bmatrix}$\\ \hline
			$y_{T+5}$ & $\begin{bmatrix}
				93.9\\ 94.6\\ 94.2
			\end{bmatrix}$ &$\begin{bmatrix}
				94.0\\ 94.2\\ 94.7
			\end{bmatrix}$ &$\begin{bmatrix}
				94.9\\ 95.3\\ 95.1
			\end{bmatrix}$ & $\begin{bmatrix}
				95.3\\ 94.4\\ 94.4
			\end{bmatrix}$ & $\begin{bmatrix}
				93.9\\ 94.5\\ 95.1
			\end{bmatrix}$ & $\begin{bmatrix}
				94.9\\ 94.4\\ 95.4
			\end{bmatrix}$\\ \hline
			$y_{T+6}$ & $\begin{bmatrix}
				94.7\\ 95.2\\ 94.6
			\end{bmatrix}$ &$\begin{bmatrix}
				95.0\\ 93.4\\ 95.0
			\end{bmatrix}$ &$\begin{bmatrix}
				94.9\\ 95.3\\ 95.1
			\end{bmatrix}$ & $\begin{bmatrix}
				94.7\\ 95.7\\ 94.5
			\end{bmatrix}$ & $\begin{bmatrix}
				95.9\\ 92.8\\ 93.7
			\end{bmatrix}$ & $\begin{bmatrix}
				94.7\\ 93.9\\ 93.4
			\end{bmatrix}$\\ \hline
			$y_{T+7}$ & $\begin{bmatrix}
				94.1\\ 94.5\\ 94.1
			\end{bmatrix}$ &$\begin{bmatrix}
				94.2\\ 95.2\\ 94.9
			\end{bmatrix}$ &$\begin{bmatrix}
				94.9\\ 95.3\\ 95.1
			\end{bmatrix}$ & $\begin{bmatrix}
				95.5\\ 94.1\\ 95.5
			\end{bmatrix}$ & $\begin{bmatrix}
				94.4\\ 94.4\\ 93.1
			\end{bmatrix}$ & $\begin{bmatrix}
				94.4\\ 95.9\\ 94.3
			\end{bmatrix}$\\ \hline
			$y_{T+8}$ & $\begin{bmatrix}
				93.4\\ 94.9\\ 94.1
			\end{bmatrix}$ &$\begin{bmatrix}
				94.0\\ 94.3\\ 95.7
			\end{bmatrix}$ &$\begin{bmatrix}
				94.9\\ 95.3\\ 95.1
			\end{bmatrix}$ & $\begin{bmatrix}
				95.4\\ 95.7\\ 95.4
			\end{bmatrix}$ & $\begin{bmatrix}
				94.5\\ 93.3\\ 93.8
			\end{bmatrix}$ & $\begin{bmatrix}
				94.8\\ 94.5\\ 94.4
			\end{bmatrix}$\\ \hline
			$y_{T+9}$ & $\begin{bmatrix}
				93.1\\ 94.6\\ 94.7
			\end{bmatrix}$ &$\begin{bmatrix}
				93.1\\ 94.8\\ 95.1
			\end{bmatrix}$ &$\begin{bmatrix}
				95.0\\ 95.0 \\95.1
			\end{bmatrix}$ & $\begin{bmatrix}
				95.0\\ 95.0\\ 95.1
			\end{bmatrix}$ & $\begin{bmatrix}
				93.1\\ 95.4\\ 93.8
			\end{bmatrix}$ & $\begin{bmatrix}
				95.8\\ 93.6\\ 93.4
			\end{bmatrix}$\\ \hline
			$y_{T+10}$ & $\begin{bmatrix}
				93.6\\ 94.0\\ 94.2
			\end{bmatrix}$ &$\begin{bmatrix}
				95.1\\ 95.3\\ 94.5
			\end{bmatrix}$ &$\begin{bmatrix}
				96.0\\ 95.1\\ 95.0
			\end{bmatrix}$ & $\begin{bmatrix}
				95.1\\ 94.6\\ 94.1
			\end{bmatrix}$ & $\begin{bmatrix}
				93.6\\ 92.5\\ 93.7
			\end{bmatrix}$ & $\begin{bmatrix}
				93.4\\ 94.4\\ 94.4
			\end{bmatrix}$\\ \hline
	\end{tabular} }
\end{table}

											\begin{table}[!ht]
												\centering
												\caption{Percentage of times that the true values lie on $95\%$ credible intervals considering $M_2$ with sample size $1000$.}\label{FORECIM21000}
											\resizebox{\columnwidth}{!}{
													\begin{tabular}{c|c|c|c|c|c|c}\hline       
														&  \multicolumn{6}{c}{Distribution of the noise process} \\\cline{2-7}
														step ahead & \multirow{2}{*}{Gaussian}&  \multirow{2}{*}{Student-$t$($\nu\!=\!5$)} & \multirow{2}{*}{Slash($\nu\!=\!4$)} & Contaminated & Symmetric & \multirow{2}{*}{Laplace} \\ 
														& & &  & normal      & hyperbolic& \\
														& & &  & ($\nu_1\!=\!0.08,\nu_2\!=\!0.012$)& ($\nu\!=\!0.12$)          & \\\hline
														$y_{T+1}$ & $\begin{bmatrix}
															95.0\\ 95.1
														\end{bmatrix}$ &$\begin{bmatrix}
															94.8\\ 94.3
														\end{bmatrix}$ &$\begin{bmatrix}
															94.9\\ 95.9
														\end{bmatrix}$ & $\begin{bmatrix}
															96.1\\ 95.9
														\end{bmatrix}$ & $\begin{bmatrix}
															94.7\\ 94.5
														\end{bmatrix}$ & $\begin{bmatrix}
															93.5\\ 93.5
														\end{bmatrix}$\\ \hline
														$y_{T+2}$ & $\begin{bmatrix}
															94.9\\ 95.3
														\end{bmatrix}$ &$\begin{bmatrix}
															96.0\\ 94.8
														\end{bmatrix}$ &$\begin{bmatrix}
															96.5\\ 95.9
														\end{bmatrix}$ & $\begin{bmatrix}
															96.4\\ 94.6
														\end{bmatrix}$ & $\begin{bmatrix}
															95.3\\ 94.6
														\end{bmatrix}$ & $\begin{bmatrix}
															94.3\\ 94.9
														\end{bmatrix}$\\ \hline
														$y_{T+3}$ & $\begin{bmatrix}
															94.9\\ 95.2
														\end{bmatrix}$ &$\begin{bmatrix}
															94.8\\ 95.5
														\end{bmatrix}$ &$\begin{bmatrix}
															95.2\\ 95.0
														\end{bmatrix}$ & $\begin{bmatrix}
															95.7\\ 95.2
														\end{bmatrix}$ & $\begin{bmatrix}
															95.4\\ 95.4
														\end{bmatrix}$ & $\begin{bmatrix}
															94.7\\ 94.9
														\end{bmatrix}$\\ \hline
														$y_{T+4}$ & $\begin{bmatrix}
															95.3\\ 94.8
														\end{bmatrix}$ &$\begin{bmatrix}
															95.2\\ 95.7
														\end{bmatrix}$ &$\begin{bmatrix}
															94.4\\ 95.8
														\end{bmatrix}$ & $\begin{bmatrix}
															96.1\\ 95.3
														\end{bmatrix}$ & $\begin{bmatrix}
															93.6\\ 95.9
														\end{bmatrix}$ & $\begin{bmatrix}
															94.1\\ 94.1
														\end{bmatrix}$\\ \hline
														$y_{T+5}$ & $\begin{bmatrix}
															95.0\\ 96.0
														\end{bmatrix}$ &$\begin{bmatrix}
															94.7\\ 94.7
														\end{bmatrix}$ &$\begin{bmatrix}
															94.3\\ 94.9
														\end{bmatrix}$ & $\begin{bmatrix}
															96.2\\ 95.8
														\end{bmatrix}$ & $\begin{bmatrix}
															93.2\\ 95.8
														\end{bmatrix}$ & $\begin{bmatrix}
															94.4\\ 94.4
														\end{bmatrix}$\\ \hline
														$y_{T+6}$ & $\begin{bmatrix}
															95.2\\ 95.1
														\end{bmatrix}$ &$\begin{bmatrix}
															95.7\\ 95.4
														\end{bmatrix}$ &$\begin{bmatrix}
															95.8\\ 94.4
														\end{bmatrix}$ & $\begin{bmatrix}
															96.3 \\96.0
														\end{bmatrix}$ & $\begin{bmatrix}
															94.0\\ 95.8
														\end{bmatrix}$ & $\begin{bmatrix}
															94.9\\ 95.8
														\end{bmatrix}$\\ \hline
														$y_{T+7}$ & $\begin{bmatrix}
															94.4 \\95.9
														\end{bmatrix}$ &$\begin{bmatrix}
															96.1\\ 96.4
														\end{bmatrix}$ &$\begin{bmatrix}
															96.6\\ 95.8
														\end{bmatrix}$ & $\begin{bmatrix}
															94.8\\ 96.4
														\end{bmatrix}$ & $\begin{bmatrix}
															94.7\\ 94.6
														\end{bmatrix}$ & $\begin{bmatrix}
															94.1\\ 94.7
														\end{bmatrix}$\\ \hline
														$y_{T+8}$ & $\begin{bmatrix}
															94.5\\ 95.2
														\end{bmatrix}$ &$\begin{bmatrix}
															95.1\\ 96.1
														\end{bmatrix}$ &$\begin{bmatrix}
															96.5\\ 95.6
														\end{bmatrix}$ & $\begin{bmatrix}
															94.2\\ 96.4
														\end{bmatrix}$ & $\begin{bmatrix}
															94.3\\ 94.5
														\end{bmatrix}$ & $\begin{bmatrix}
															93.7\\ 95.0
														\end{bmatrix}$\\ \hline
														$y_{T+9}$ & $\begin{bmatrix}
															95.1\\ 94.7
														\end{bmatrix}$ &$\begin{bmatrix}
															96.0\\ 96.9
														\end{bmatrix}$ &$\begin{bmatrix}
															95.9\\ 94.5
														\end{bmatrix}$ & $\begin{bmatrix}
															95.3\\ 95.6
														\end{bmatrix}$ & $\begin{bmatrix}
															94.5\\ 94.2
														\end{bmatrix}$ & $\begin{bmatrix}
															94.1\\ 95.2
														\end{bmatrix}$\\ \hline
														$y_{T+10}$ & $\begin{bmatrix}
															94.9\\ 95.0
														\end{bmatrix}$ &$\begin{bmatrix}
															94.0\\ 94.5
														\end{bmatrix}$ &$\begin{bmatrix}
															95.4\\ 95.8
														\end{bmatrix}$ & $\begin{bmatrix}
															93.8 \\95.1
														\end{bmatrix}$ & $\begin{bmatrix}
															95.8\\ 94.1
														\end{bmatrix}$ & $\begin{bmatrix}
															94.2\\ 93.4
														\end{bmatrix}$\\ \hline
												\end{tabular} }
											\end{table}

																						\clearpage
\subsection{Extra Simulation for Regimes and Orders}
\label{subsec43}
We also carried out an extra simulation to check if the DAIC and WAIC criteria can be used to identify the autoregressive orders in each regime, the number of regimes, and the distribution of errors because the proposed methodology is strongly dependent on those three aspects. We focus this empirical exploration on simulating 1000 time series from MTAR models $M_1$ and $M_2$ with the same parameters but with only autoregressive $p=(1,1)$ or $p=(1,1,1)$, respectively, and a distribution error that follows a multivariate student-t with 5 degrees of freedom.

We initially calculated the proportion of times that DAIC and WAIC select a model with the correct autoregressive orders in each regime, in contrast to models of autoregressive orders 2 and 3 with the same number of regimes. Table(\ref{ProportionM11000Orders}) shows the results for time series of length $T=1000$. DIC a WAIC works reasonably well in correctly determining the autoregressive orders for two-regimes models because the proportion is high for both criteria; however, the performance is decreased when the number of regimes is increased.

\begin{table}[!ht]
	\centering
	\caption{Proportion of times that DIC and WAIC select the true autoregressive order in each regime using sample size $T=1000$.}\label{ProportionM11000Orders}
	{\footnotesize
		\begin{tabular}{c|c|c|}\hline       
			&  \multicolumn{2}{c}{True Model} \\\cline{2-3}
			Criteria&   $M_1$ with $p=(1,1)$ & $M_2$ with $p=(1,1,1)$  \\\hline
			DIC  & 0.881 & 0.569    \\
			WAIC & 0.86 & 0.592   \\ \hline
	\end{tabular}}
\end{table}

We also computed the proportion of times that DAIC and WAIC select a model with the correct number of regimes in contrast with models linear VAR, MTAR with 3 and 4 regimes; all of them were adjusted with parameter autoregressive 1 in each regime. Table (\ref{ProportionM11000Regimes}) shows that this proportion is high when the number of regimes is two and is decreased when the number of regimes increases to 3. Additionally, we can observe that both criteria work similarly.




\begin{table}[!ht]
	\centering
	\caption{Proportion of times that DIC and WAIC select the true number of regimes using sample size $T=1000$.}\label{ProportionM11000Regimes}
	{\footnotesize
		\begin{tabular}{c|c|c|}\hline       
			&  \multicolumn{2}{c}{True Model} \\\cline{2-3}
			Criteria&   2 Regimes & 3 Regimes with Student-t distribution   \\\hline
			DIC &  0.9361 & 0.639    \\
			WAIC&  0.947  & 0.652   \\ \hline
	\end{tabular}}
\end{table}







Finally, we also conducted an experiment to check whether DIC and WAIC selected the error distribution appropriately. In Table(\ref{ProportionM11000Dist}), we can see the proportion of times that the criteria correctly select the true distribution or, in parentheses, the proportion of times that criteria select the true distribution in the second place. We can observe that WAIC performs better than DIC, in particular for models with two regimes; furthermore, the proportion is dramatically decreased when the number of regimes of the true model is increased to 3.  However, it is interesting to consider that the true distribution of errors can be found if the criterion is checked for the first two lower values.

\begin{table}[!ht]
	\centering
	\caption{Proportion of times that DIC and WAIC select the true distribution using sample size $T=1000$.}\label{ProportionM11000Dist}
\resizebox{\columnwidth}{!}{
		\begin{tabular}{c|c|c|}\hline       
			&  \multicolumn{2}{c}{True Model} \\\cline{2-3}
			Criteria&   2 Regimes with Student-t distribution & 3 Regimes  with Student-t distribution   \\\hline
			DIC & 0.030(0.693)  & 0.163(0.710)    \\
			WAIC& 0.720(0.223)  & 0.186(0.740)   \\ \hline
	\end{tabular}}
\end{table}













\clearpage
\section{Real Data Application}
\label{sec5}
This section analyzes two real multivariate time series to illustrate the proposed methodology. The {\tt R} package {\tt mtarm} is used here, as it includes both the time series data to be analyzed and the routines in which the proposed procedures for estimating and forecasting multivariate TAR models are implemented. We selected the hyperparameters of the prior distributions in order to obtain non-informative distributions.

\subsection{Rainfall and two river flows in Colombia}
\label{subsec51}
This time series was also considered in \cite{CN17} in the context of hydrological/meteorological daily time series by analyzing the relationship between daily rainfall (in $mm$) and two river flows (in $m^3\!/\!s$) from 2006-01-01 to 2009-04-14 (1200 time points). Rainfall was measured at the San Juan meteorological station, the flow of the Bedon river was determined at the El Trebol hydrological station, and the flow of the La Plata river was measured at the Villalosada hydrological station. The missing data were replaced by the results of the procedure implemented by \cite{CN17}. Thus, $Y_t$ is a bivariate time series comprised of the flow of the Bedon and La Plata rivers, whereas $Z_t$ is the rainfall time series. These data are available in the object {\tt riverflows} of the package {\tt mtarm}. The last ten observations were left out to assess the forecasting procedure. \\

Table \ref{Criteriaregimesriverflowexample} presents the values of the DIC and WAIC criteria for ninety different models: ${\rm MTAR}(3;p=(p^*,p^*,p^*)^{\!\top})$, ${\rm MTAR}(2;p=(p^*,p^*)^{\!\top})$, and ${\rm VAR}(p^*)$ for $p^*=1,\ldots,5$ and all the distributions described above. The TAR-type nonlinearity is first assessed. The values of the criteria DIC and WAIC suggest that the nonlinear models ${\rm MTAR}(2)$ and ${\rm MTAR}(3)$ are better than the linear ones regardless of the value of $p^*$ and the distribution used to describe the noise process. The values of $l$ and $p^*$ suggested by the criteria DIC and WAIC are $3$ and $5$, respectively. In addition, the Laplace distribution, whose tails are ``rather heavier'' than those of the Gaussian distribution is chosen to describe the noise process simply because it presents the lowest values on the DIC and WAIC criteria. Therefore, the chosen model is ${\rm MTAR}(3;p=(5,5,5)^{\!\top})$, where the noise process is Laplace distributed.


\begin{table}[!ht]
	\centering
	\caption{DIC and WAIC values of some models fitted to river flow time series. In bold, the lowest values for each criterion and distribution. $\ast$ and $\ast\ast$ mean first and second lowest values, respectively, in each criterion.}
	\label{Criteriaregimesriverflowexample}
	\resizebox{\columnwidth}{!}{
		\begin{tabular}{c|c|c|c|c|c|c|c|c}   \hline  
			\multirow{3}{*}{Criteria} & \multirow{3}{*}{Model}& \multirow{3}{*}{$p^*$} & \multicolumn{6}{c}{Distribution of the noise process} \\\cline{4-9}
			&                           &     & \multirow{2}{*}{Gaussian} & \multirow{2}{*}{Student-$t$} & \multirow{2}{*}{Slash} & Contaminated & Symmetric & \multirow{2}{*}{Laplace}\\ 
			&                           &     &                           &                              &                        &  normal      & hyperbolic& \\\hline
			\multirow{15}{*}{DIC} & \multirow{5}{*}{VAR}	  &  1  &  15399.13 & 14474.11 &   14533.31 &   14503.73 &  14510.15  &	  14490.85 \\              
			& 				       	  &  2	&  15335.70 & 14412.83 &   14469.90 &	14448.68 &  14447.22  &   14423.03 \\        
			& 				       	  &  3	&  15311.62 & 14390.58 &   14448.68 &	14429.53 &  14425.44  &   14399.62 \\        
			& 				       	  &  4	&  15293.80 & 14375.99 &   14432.67 &	14413.16 &  14407.06  &   14379.55 \\        
			& 				       	  &  5	&  15285.04 & 14366.57 &   14422.70 &	14403.04 &  14401.52  &   14373.25 \\\cline{2-9}  
			& \multirow{5}{*}{MTAR(2)}  &  1	&  14120.27 & 13790.51 &   13721.82 &	13790.75 &  13747.48  &   13757.28 \\        
			& 				       	  &  2  &  14057.27 & 13675.44 &   13681.03 &	13746.99 &  13713.60  &   13706.83 \\        
			& 				       	  &  3  &  14007.15 & 13703.40 &   13630.51 &	13707.19 &  13686.49  &   13680.01 \\        
			& 				       	  &  4  &  13997.64 & 13688.65 &   13619.51 &	13690.52 &  13664.21  &   13656.17 \\        
			& 				       	  &  5  &  13979.47 & 13687.49 &   13725.01 &	13683.95 &  13655.03  &   13644.09 \\\cline{2-9} 
			& \multirow{5}{*}{MTAR(3)}  &  1	&  13987.89 & 13589.89 &   13625.59 &	13596.77 &  13574.04  &   13572.24 \\        
			& 				       	  &  2  &  13951.73 & 13553.70 &   13585.81 &	13559.81 &  13540.08  &   13534.62 \\        
			& 				       	  &  3  &  13850.39 & 13531.00 &   13561.60 &	13536.67 &  13519.64  &   13515.78 \\        
			& 				       	  &  4  &  13895.55 & 13513.14 &   13542.56 &	13519.57 &  13501.05  &   13496.18 \\        
			& 				       	  &  5  &  {\bf13835.36} & {\bf13500.64} &   {\bf13533.51} &   {\bf13507.79} &  \,\,\,\,\,{\bf13477.67}$\ast\ast$  &   \,\,{\bf13468.70}$\ast$ \\\hline  
			\multirow{15}{*}{WAIC}& \multirow{5}{*}{VAR}	  &  1	&  15405.49 & 14476.39 &   14537.12 &	14511.48 &  14514.21  &   14494.69 \\      		       			    					  & 				       	  &  2  &  15343.24 & 14417.69 &   14475.65 &	14459.06 &  14451.87  &   14428.41 \\      	       			    
			& 				       	  &  3  &  15319.15 & 14395.97 &   14456.31 &	14442.83 &  14429.69  &   14406.08 \\      	       			    
			& 				       	  &  4  &  15302.12 & 14383.04 &   14441.32 &	14428.14 &  14413.30  &   14387.01 \\      	       			    
			& 				       	  &  5  &  15293.92 & 14375.18 &   14433.10 &	14419.11 &  14407.47  &   14379.82 \\\cline{2-9}	       			    
			& \multirow{5}{*}{MTAR(2)}  &  1	&  14131.24 & 13795.56 &   13727.34 &	13800.82 &  13769.07  &   13765.14 \\      	       			    
			& 				       	  &  2  &  14071.34 & 13683.18 &   13689.33 &	13760.96 &  13723.29  &   13720.48 \\      	       			    
			& 				       	  &  3  &  14021.86 & 13714.33 &   13640.39 &	13723.51 &  13704.62  &   13703.37 \\      	       			    
			& 				       	  &  4  &  14014.13 & 13701.09 &   13631.00 &	13708.80 &  13682.09  &   13680.56 \\      	       			    
			& 				       	  &  5  &  13995.68 & 13702.87 &   13741.07 &	13707.67 &  13673.79  &   13667.27 \\\cline{2-9}       			    
			& \multirow{5}{*}{MTAR(3)}  &  1	&  14045.06 & 13597.06 &   13634.19 &	13607.17 &  13586.44  &   13585.02 \\      	       			    
			& 				       	  &  2  &  13973.44 & 13567.42 &   13601.42 &	13576.41 &  13559.03  &   13558.54 \\      	       			    
			& 				       	  &  3  &  13975.36 & 13549.29 &   13581.25 &	13558.71 &  13543.81  &   13545.55 \\      	       			    
			& 				       	  &  4  &  13918.88 & 13534.09 &   13564.69 &	13545.47 &  13532.25  &   13534.93 \\      	       			    
			& 				       	  &  5  &  {\bf13901.10} & {\bf13525.69} &   {\bf13558.77} &	{\bf13544.79} &  \,\,\,\,\,{\bf13515.57}$\ast\ast$  &   \,\,{\bf13514.41}$\ast$ \\\hline
	\end{tabular}}
\end{table}

	
	\begin{table}
		\centering
		\caption{Posterior mean of the parameters in the ${\rm MTAR}(3;p=(5,5,5)^{\!\top})$ model where the noise process is Laplace distributed when fitted to the river flow time series. 
		}
		\label{Estimationriverflowexample}
	\resizebox{\columnwidth}{!}{
			\begin{tabular}{c|c|c|c} 
				Parameter &  Regime $j=1$ & Regime $j=2$ & Regime $j=3$ \\\hline
				${\phi}_0^{(j)}$ & $\begin{bmatrix}       
					1.30796\\
					3.48465\\
				\end{bmatrix}$   & $\begin{bmatrix}
					2.09284\\
					6.73924\\
				\end{bmatrix}$  & $\begin{bmatrix}
					5.82170\\
					18.56190\\
				\end{bmatrix}$ \\
				$\bm{\phi}_1^{(j)}$ & $\begin{bmatrix}
					0.56647 & 0.04427\\
					0.17070 & 0.61034\\
				\end{bmatrix}$ &   $\begin{bmatrix}
					0.58561 & 0.02149\\
					0.14502 & 0.53452\\
				\end{bmatrix}$ & $\begin{bmatrix}
					0.44201 & 0.04376\\
					0.43390 & 0.33443\\
				\end{bmatrix}$\\
				
				$\bm{\phi}_2^{(j)}$& $\begin{bmatrix}
					0.04929 & -0.01868\\
					-0.05533 & -0.05392\\
				\end{bmatrix}$ & $\begin{bmatrix}
					0.09018 & -0.02082\\
					-0.01639 &  0.02130\\
				\end{bmatrix}$  & $\begin{bmatrix}
					0.11482 & -0.00362\\
					-0.47808 &  0.10884\\
				\end{bmatrix}$\\
				
				$\bm{\phi}_3^{(j)}$& $\begin{bmatrix}
					0.02023 & 0.00702\\
					0.02423 & 0.07010\\
				\end{bmatrix}$ &  $\begin{bmatrix}
					-0.03170 & -0.00651\\
					-0.04980 &  0.04530\\
				\end{bmatrix}$ & $\begin{bmatrix}
					-0.10619 & 0.03434\\
					-0.60310 & 0.28298\\
				\end{bmatrix}$\\
				
				$\bm{\phi}_4^{(j)}$& $\begin{bmatrix}
					0.03685 & -0.01594\\
					-0.07863 & 0.00042\\
				\end{bmatrix}$ & $\begin{bmatrix}
					0.10739 & 0.00691\\
					0.23162 & -0.03936\\
				\end{bmatrix}$ & $\begin{bmatrix}
					0.00145 & -0.00054\\
					0.05003 & -0.02254\\
				\end{bmatrix}$\\
				
				$\bm{\phi}_5^{(j)}$& $\begin{bmatrix}
					0.08331 & -0.00766\\
					0.13931 &  0.02996\\
				\end{bmatrix}$ &  $\begin{bmatrix}
					0.02949 & 0.00330\\
					-0.25938 & 0.11347\\
				\end{bmatrix}$ & $\begin{bmatrix}
					0.17848 & -0.00917\\
					0.24315 &  0.08332\\
				\end{bmatrix}$\\
				
				$\bm{\Sigma}_j$ & $\begin{bmatrix}
					0.32786  &  0.36473\\
					0.36473  &  2.34362 \\
				\end{bmatrix}$ & $\begin{bmatrix}
					1.06882  &  1.27780\\
					1.27780  &  6.38990\\
				\end{bmatrix}$ & $\begin{bmatrix}
					2.81492  &  7.23904\\
					7.23904  & 43.13993\\
				\end{bmatrix}$  \\ \hline 
				$c_1$  & \multicolumn{3}{l}{\qquad\qquad$3$}\\
				$c_2$  & \multicolumn{3}{l}{\qquad\qquad$10$}\\
				$h$ & \multicolumn{3}{l}{\qquad\qquad$0$} \\ \hline
		\end{tabular}}
	\end{table}
	
	Table \ref{Estimationriverflowexample} presents the posterior means of the parameters in the chosen model. That model allows to conclude the following: $(i)$ three regimes of precipitation influence river flows: a regime of few precipitations (less than 3 $mm$) and low river flow, a second regime of intermediate precipitation (between 3 and 10 $mm$) and medium river flow, and a third regime of high precipitation (more than 10 $mm$) and high river flow; $(ii)$ the river flows depend on the current precipitation level, as the posterior mean of the delay parameter $h$ is $0$; and $(iii)$ the variability increases with the regime, as the diagonal values in the posterior means of $\bm{\Sigma}_1$ are lower than those of $\bm{\Sigma}_2$, and the latter are lower than those of $\bm{\Sigma}_3$. Unlike the ${\rm MTAR}(3;p=(5,5,5)^{\!\top})$ model where the noise process is Gaussian distributed, the normal QQ-plot of $r_t$ for the ${\rm MTAR}(3;p=(5,5,5)^{\!\top})$ model where the noise process is Laplace distributed (not shown here) is close to a line with unit slope and zero intercept, thus indicating a suitable model fit. The chosen model and the real values of rainfall are used to obtain a forecast of 10 steps ahead of the Bedon and La Plata river flows (see Table \ref{Forecastingriverflowexample}). The forecasts are close to their true values for all steps ahead. Even though the Laplace distribution can generate values across the entire real line, many prediction intervals include only positive values.
	
	\begin{table}
		\centering
		\caption{Forecast of 10 steps-ahead for the river flows time series from the fitted ${\rm MTAR}(3;p=(5,5,5)^{\!\top})$ model where the noise process is Laplace distributed.}
		\label{Forecastingriverflowexample}
		{\footnotesize
			\begin{tabular}{c|c|c|c} 
				step-ahead &  True value  & Forecast & $95\%$ prediction interval\\\hline
				\multirow{2}{*}{1} & $\begin{bmatrix}
					\,\,9.456     \\    19.600
				\end{bmatrix}$ & $\begin{bmatrix}
					\,\,9.705 \\
					24.189\\
				\end{bmatrix}$
				&
				$\begin{bmatrix}
					(\,6.147,\,12.958)\\        
					(14.799,\,33.674)
				\end{bmatrix}$\\\hline
				\multirow{2}{*}{2} & $\begin{bmatrix}
					10.400  \\         25.260
				\end{bmatrix}$ & $\begin{bmatrix}
					13.902      \\
					34.995\\
				\end{bmatrix}$
				&
				$\begin{bmatrix}
					(\,\,\,3.584,\,22.895)\\        
					(-5.279,\,73.113)
				\end{bmatrix}$\\\hline
				\multirow{2}{*}{3} & $\begin{bmatrix}
					13.240   \\         42.280 
				\end{bmatrix}$ & $\begin{bmatrix}
					16.306      \\
					41.169\\
				\end{bmatrix}$
				&
				$\begin{bmatrix}
					(4.803,\,27.136)\\         
					(2.648,\,85.188)
				\end{bmatrix}$\\
				\hline
				\multirow{2}{*}{4} & $\begin{bmatrix}
					11.360    \\        33.450
				\end{bmatrix}$ & $\begin{bmatrix}
					14.279  \\
					33.494\\
				\end{bmatrix}$
				&
				$\begin{bmatrix}
					(5.089,\,24.384)\\         
					(6.148,\,62.089)
				\end{bmatrix}$\\\hline
				\multirow{2}{*}{5} & $\begin{bmatrix}
					11.170    \\        32.400
				\end{bmatrix}$ & $\begin{bmatrix}
					12.760   \\
					30.118\\
				\end{bmatrix}$
				&
				$\begin{bmatrix}
					(4.159,\,21.343)\\         
					(7.943,\,53.047)
				\end{bmatrix}$\\\hline
				\multirow{2}{*}{6} & $\begin{bmatrix}
					10.590     \\       25.170 
				\end{bmatrix}$ & $\begin{bmatrix}
					12.131   \\
					28.391\\
				\end{bmatrix}$
				&
				$\begin{bmatrix}
					(3.770,\,20.238)\\         
					(7.516,\,47.814)
				\end{bmatrix}$\\
				\hline
				\multirow{2}{*}{7} & $\begin{bmatrix}
					11.460  \\          42.730
				\end{bmatrix}$ & $\begin{bmatrix}
					15.574   \\
					37.940\\
				\end{bmatrix}$
				&
				$\begin{bmatrix}
					(\,\,\,4.412,\,27.796)\\        
					(-2.156,\,79.589)
				\end{bmatrix}$\\\hline
				\multirow{2}{*}{8} & $\begin{bmatrix}
					12.447        \\       29.270
				\end{bmatrix}$ & $\begin{bmatrix}
					14.276   \\
					32.708\\
				\end{bmatrix}$
				&
				$\begin{bmatrix}
					(4.906,\,24.312)\\         
					(3.871,\,60.132)
				\end{bmatrix}$\\
				\hline
				\multirow{2}{*}{9} & $\begin{bmatrix}
					11.460    \\        27.380 
				\end{bmatrix}$ & $\begin{bmatrix}
					12.223  \\
					27.219\\
				\end{bmatrix}$
				&
				$\begin{bmatrix}
					(4.138,\,19.057)\\         
					(8.363,\,46.418)
				\end{bmatrix}$\\\hline
				\multirow{2}{*}{10} & $\begin{bmatrix}
					11.070    \\        25.820 
				\end{bmatrix}$ & $\begin{bmatrix}
					10.880   \\
					24.313
				\end{bmatrix}$
				&
				$\begin{bmatrix}
					(4.356,\,16.993)\\         
					(7.000,\,39.295)
				\end{bmatrix}$
		\end{tabular}}
	\end{table}

	\subsection{Returns of the closing prices of three financial indexes}
	
	This time series was analyzed by \cite{RC21} and corresponded to the returns of closing prices of the Colcap, Bovespa, and S\&P 500 indexes from 2010-02-01 to 2016-03-31 (1505 time points). Colcap is a leading indicator of the price dynamics of the 20 most liquid shares on the Colombian Stock Market. Bovespa is the Brazilian stock market index, the world's thirteenth largest and most important stock exchange, and the first in Latin America. Finally, the Standard \& Poor's 500 (S\&P 500) index is a stock index based on the 500 largest companies in the United States. Thus, $Y_t$ is a bivariate time series comprised of the returns of the Colcap and Bovespa indexes, whereas $Z_t$ is the time series of the returns of the S\&P 500 index. These data are available in the object {\tt returns} of the package {\tt mtarm}. The last ten observations were left out to assess the forecasting procedure.\\
	
	
	Table \ref{Criteriaregimesfinancialexample} presents the values of the DIC and WAIC criteria for ninety different models: ${\rm MTAR}(3;p=(p^*,p^*,p^*)^{\!\top})$, ${\rm MTAR}(2;p=(p^*,p^*)^{\!\top})$, and ${\rm VAR}(p^*)$ for $p^*=1,\ldots,5$ and all the distributions described above.
	The values of the criteria DIC and WAIC suggest that the nonlinear models ${\rm MTAR}(2)$ and ${\rm MTAR}(3)$ are better than the linear ones regardless
	of the value of $p^*$ and the distribution used to describe the noise process. The chosen model is ${\rm MTAR}(3;p=(2,2,2)^{\!\top})$, where the noise
	process is Student-$t$ distributed, as its value on the WAIC criterion is the lowest.
	
	\begin{table}[!ht]
		\centering
		\caption{DIC and WAIC values of some models fitted the time series of returns of the indexes Colcap and Bovespa. In bold, the lowest values for each criterion and distribution. $\ast$ and $\ast\ast$ mean first and second lowest values, respectively, in each criterion.}
		\label{Criteriaregimesfinancialexample}
		\resizebox{\columnwidth}{!}{
			\begin{tabular}{c|c|c|c|c|c|c|c|c}   \hline  
				\multirow{3}{*}{Criteria} & \multirow{3}{*}{Model}& \multirow{3}{*}{$p^*$} & \multicolumn{6}{c}{Distribution of the noise process} \\\cline{4-9}
				&                           &     & \multirow{2}{*}{Gaussian} & \multirow{2}{*}{Student-$t$} & \multirow{2}{*}{Slash} & Contaminated & Symmetric & \multirow{2}{*}{Laplace}\\ 
				&                           &     &                           &                              &                        &  normal      & hyperbolic& \\\hline
				\multirow{15}{*}{DIC} & \multirow{5}{*}{VAR}	    &  1  &  -18325.97 & -18499.04 &   -18470.52 &    -18504.39 &  -18518.66  &	-18481.56 \\              
				& 				       	  &  2	&  -18322.17 & -18496.21 &   -18467.18 &	-18501.49 &  -18713.56  &   -18488.92 \\        
				& 				       	  &  3	&  -18318.17 & -18491.96 &   -18463.28 &	-18492.78 &  -18717.93  &   -18488.37 \\        
				& 				       	  &  4	&  -18321.22 & -18490.31 &   -18462.44 &	-18489.50 &  -18748.48  &   -18481.65 \\        
				& 				       	  &  5	&  -18319.98 & -18486.55 &   -18459.54 &	-18495.00 &  -18933.48  &   -18473.90 \\\cline{2-9}  
				& \multirow{5}{*}{MTAR(2)}  &  1	&  -18623.13 & -18764.07 &   -18758.78 &	-18768.99 &  -18779.49  &   -18655.96 \\        
				& 				       	  &  2  &  -18623.88 & -18760.58 &   -18739.32 &	-18738.22 &  -19101.64  &   -18663.00 \\        
				& 				       	  &  3  &  -18613.92 & -18751.62 &   -18734.36 &	-18727.86 &  -18993.81  &   -18683.19 \\        
				& 				       	  &  4  &  -18610.80 & -18723.50 &   -18714.37 &	-18736.34 &  -18793.65  &   -18676.65 \\        
				& 				       	  &  5  &  -18610.94 & -18743.51 &   -18717.50 &	-18716.28 &  -18760.43  &   -18658.28 \\\cline{2-9}  
				& \multirow{5}{*}{MTAR(3)}  &  1	&  -18872.79 & -18969.75 &   -18958.07 &	-18971.83 &  -19021.75  &   -18785.23 \\        
				& 				       	  &  2  &  -18875.53 &{\bf-19008.57}&{\bf-18981.35}&-18887.62 &\,\,\,\,{\bf-19243.62}$\ast\ast$& -18822.78 \\        
				& 				       	  &  3  &  -18876.13 & -18965.87 &   -18914.41 &\,\,{\bf-19354.00}$\ast$&  -18841.83  &  {\bf-18824.92}\\        
				& 				       	  &  4  &{\bf-18891.02}& -18950.77 & -18936.17 &	-18819.76 &  -18802.23  &   -18818.50 \\        
				& 				       	  &  5  &  -18869.07 & -18955.67 &   -18956.29 &    -18851.56 &  -18800.40  &   -18815.76 \\\hline  
				\multirow{15}{*}{WAIC}& \multirow{5}{*}{VAR}	    &  1  &  -18322.50 & -18498.48 &   -18469.74 &	-18501.63 &  -18515.59  &   -18480.74 \\      		       	 					  & 				       	  &  2  &  -18317.45 & -18495.02 &   -18465.64 &	-18497.16 &  -18509.88  &   -18486.73 \\      	       		 					  & 				       	  &  3  &  -18311.90 & -18489.98 &   -18460.68 &	-18486.41 &  -18496.48  &   -18484.23 \\      	       		 					  & 				       	  &  4  &  -18314.54 & -18487.91 &   -18459.65 &	-18481.92 &  -18483.67  &   -18474.82 \\      	       		 					  & 				       	  &  5  &  -18312.85 & -18483.84 &   -18456.33 &	-18477.97 &  -18475.99  &   -18464.00 \\\cline{2-9}	       		 					  & \multirow{5}{*}{MTAR(2)}  &  1	&  -18610.72 & -18745.14 &   -18730.18 &	-18746.48 &  -18757.21  &   -18645.60 \\      	       		 					  & 				       	  &  2  &  -18608.21 & -18756.35 &   -18737.54 &	-18734.45 &  -18678.97  &   -18656.29 \\      	       		 					  & 				       	  &  3  &  -18594.72 & -18711.49 &   -18688.68 &	-18718.06 &  -18678.23  &   -18644.57 \\      	       		 					  & 				       	  &  4  &  -18583.79 & -18718.07 &   -18680.66 &	-18714.15 &  -18649.46  &   -18616.66 \\      	       		 					  & 				       	  &  5  &  -18587.95 & -18736.49 &   -18703.39 &	-18705.35 &  -18640.55  &   -18638.29 \\\cline{2-9}	       		 					  & \multirow{5}{*}{MTAR(3)}  &  1	&{\bf-18855.85}& -18965.12 & -18954.44 &{\bf-18965.33}&{\bf-18833.17}&  -18777.32 \\      	       		 					  & 				       	  &  2  &  -18843.29 &\,\,{\bf-19002.79}$\ast$&\,\,\,\,{\bf-18971.94}$\ast\ast$&-18862.67 &  -18828.07  &   -18792.10 \\      	       		 					  & 				       	  &  3  &  -18819.17 & -18935.82 &   -18901.09 &	-18785.35 &  -18811.49  &{\bf-18797.32}\\      	       		 					  & 				       	  &  4  &  -18853.78 & -18941.78 &   -18912.20 &	-18755.41 &  -18783.67  &   -18790.13 \\      	       		 					  & 				       	  &  5  &  -18837.65 & -18942.89 &   -18928.59 &	-18777.18 &  -18775.67  &   -18779.85 \\\hline      	       \end{tabular}}
	\end{table}

		Table \ref{Estimationfinancialexample} presents the posterior means of the parameters in the chosen model. That model allows to conclude the following: $(i)$ the posterior mean of the delay parameter $h$ is $0$, which suggests that the effect of the S\&P 500 index is instant on the Colcap and Bovespa indexes; $(ii)$ the first regime is for losses, the second regime is a kind of equilibrium, while the third regime is related to gains for the Colcap and Bovespa indexes; $(iii)$ 
		similar variability is observed in all regimes due to the diagonal values of the posterior means of the matrices $\bm{\Sigma}_1$, $\bm{\Sigma}_2$ and $\bm{\Sigma}_3$ are similar; $(iv)$ the presence of heavy tails due to the chosen noise process distribution, which is a Student-$t$ with a ``small'' extra parameter value ($\nu=5.757$). Unlike the ${\rm MTAR}(3;p=(2,2,2)^{\!\top})$ model where the noise process is Gaussian distributed, the normal QQ-plot of $r_t$ for the ${\rm MTAR}(3;p=(2,2,2)^{\!\top})$ model where the noise process is Student-$t$ distributed (not shown here) is close to a line with unit slope and zero intercept, thus indicating a suitable model fit. The chosen model was used to obtain the forecast ten steps ahead of the indexes Colcap and Bovespa (Table \ref{Forecastingfinancialexample}), by assuming that the true values of the S\&P 500 index are known. In most cases, the $95\%$ prediction interval captures the true value.
		
		\begin{table}
			\centering
			\caption{Posterior mean of the parameters in the ${\rm MTAR}(3;p=(2,2,2)^{\!\top})$ model where the noise process is Student-$t$ distributed when fitted to the time series of returns of the indexes Colcap and Bovespa. 
			}
			\label{Estimationfinancialexample}
			\resizebox{\columnwidth}{!}{
				\begin{tabular}{c|c|c|c} 
					Parameter &  Regime $j=1$ & Regime $j=2$ & Regime $j=3$ \\\hline
					${\phi}_0^{(j)}$ & $\begin{bmatrix}       
						-0.00769\\
						-0.01447\\
					\end{bmatrix}$   & $\begin{bmatrix}
						0.00007\\
						-0.00066\\
					\end{bmatrix}$  & $\begin{bmatrix}
						0.00585 \\
						0.01274\\
					\end{bmatrix}$ \\
					$\bm{\phi}_1^{(j)}$ & $\begin{bmatrix}
						0.19510 &  0.12894\\
						-0.10091 &  0.12886\\
					\end{bmatrix}$ &   $\begin{bmatrix}
						0.05303 &  0.07932\\
						0.04873 & -0.03413\\
					\end{bmatrix}$ & $\begin{bmatrix}
						0.10868 &  0.04200\\
						0.14794 & -0.08644\\
					\end{bmatrix}$\\
					
					$\bm{\phi}_2^{(j)}$& $\begin{bmatrix}
						0.10502 &  0.04845\\
						0.08285 &  0.05409\\
					\end{bmatrix}$ & $\begin{bmatrix}
						-0.02671 &   0.04095\\
						0.10437 &  -0.03063\\
					\end{bmatrix}$  & $\begin{bmatrix}
						0.12312 & -0.11415\\
						-0.01006 & -0.08627\\
					\end{bmatrix}$\\
					$\bm{\Sigma}_j$ & $\begin{bmatrix}
						0.000081  &  0.000041\\
						0.000041  &  0.000151\\
					\end{bmatrix}$ & $\begin{bmatrix}
						0.000042  &  0.000015\\
						0.000015  &  0.000091\\
					\end{bmatrix}$ & $\begin{bmatrix}
						0.000060  &  0.000021\\
						0.000021  &  0.000142\\
					\end{bmatrix}$  \\ \hline 
					$c_1$  & \multicolumn{3}{l}{\qquad\qquad$-0.01$}\\
					$c_2$  & \multicolumn{3}{l}{\qquad\qquad$0.00942$}\\
					$h$ & \multicolumn{3}{l}{\qquad\qquad$0$} \\
					$\nu$ & \multicolumn{3}{l}{\qquad\qquad$5.757$} \\\hline
			\end{tabular}}
		\end{table}
		
		\begin{table}
			\centering
			\caption{Forecast of 10 steps-ahead for the time series of returns of the indexes Colcap and Bovespa from the fitted ${\rm MTAR}(3;p=(2,2,2)^{\!\top})$ model  where the noise process is Student-$t$ distributed.}
			\label{Forecastingfinancialexample}
			{\footnotesize
				\begin{tabular}{c|c|c|c} 
					Step-ahead &  True value  & Forecast & $95\%$ Prediction interval\\\hline
					\multirow{2}{*}{1} & $\begin{bmatrix}
						-0.010171 \\ -0.036199
					\end{bmatrix}$ & $\begin{bmatrix}
						-0.001289     \\
						-0.000109\\
					\end{bmatrix}$
					&
					$\begin{bmatrix}
						(-0.017755,\, 0.013928)\\
						(-0.023845,\, 0.023754)
					\end{bmatrix}$\\\hline
					\multirow{2}{*}{2} & $\begin{bmatrix}
						-0.000732\\ 0.013350
					\end{bmatrix}$ & $\begin{bmatrix}
						-0.000386     \\
						-0.000212\\
					\end{bmatrix}$
					&
					$\begin{bmatrix}
						(-0.015541 ,\, 0.016441)\\
						(-0.022695  ,\, 0.021623)
					\end{bmatrix}$\\\hline
					\multirow{2}{*}{3} & $\begin{bmatrix}
						0.008906   \\  0.063874
					\end{bmatrix}$ & $\begin{bmatrix}
						-0.000030       \\
						-0.000733\\
					\end{bmatrix}$
					&
					$\begin{bmatrix}
						(-0.015599 ,\,0.016283)\\
						(-0.023246  ,\, 0.022471)
					\end{bmatrix}$\\
					\hline
					\multirow{2}{*}{4} & $\begin{bmatrix}
						0.010684 \\ -0.001949
					\end{bmatrix}$ & $\begin{bmatrix}
						-0.000294         \\
						-0.001050\\
					\end{bmatrix}$
					&
					$\begin{bmatrix}
						(-0.015822 ,\, 0.016163)\\
						(-0.023661 ,\, 0.025639)
					\end{bmatrix}$\\\hline
					\multirow{2}{*}{5} & $\begin{bmatrix}
						0.005414\\  0.003841
					\end{bmatrix}$ & $\begin{bmatrix}
						-0.000405          \\
						-0.000792\\
					\end{bmatrix}$
					&
					$\begin{bmatrix}
						(-0.016190 ,\, 0.014867)\\
						(-0.024279  ,\, 0.022540)
					\end{bmatrix}$\\\hline
					\multirow{2}{*}{6} & $\begin{bmatrix}
						0.003131 \\ -0.026221
					\end{bmatrix}$ & $\begin{bmatrix}
						-0.000047      \\
						-0.000668\\
					\end{bmatrix}$
					&
					$\begin{bmatrix}
						(-0.015579 ,\, 0.016173)\\
						(-0.023675  ,\, 0.023678)
					\end{bmatrix}$\\
					\hline
					\multirow{2}{*}{7} & $\begin{bmatrix}
						-0.011910 \\ 0.022844
					\end{bmatrix}$ & $\begin{bmatrix}
						0.000067        \\
						-0.000669\\
					\end{bmatrix}$
					&
					$\begin{bmatrix}
						(-0.015560 ,\, 0.017264)\\
						(-0.024937 ,\, 0.022864)
					\end{bmatrix}$\\\hline
					\multirow{2}{*}{8} & $\begin{bmatrix}
						-0.004422   \\ 0.006211
					\end{bmatrix}$ & $\begin{bmatrix}
						-0.000029       \\
						-0.000627\\
					\end{bmatrix}$   
					&
					$\begin{bmatrix}
						(-0.017748,\, 0.014119)\\
						(-0.024352  ,\, 0.024552)
					\end{bmatrix}$\\
					\hline
					\multirow{2}{*}{9} & $\begin{bmatrix}
						0.015362   \\ 0.001835
					\end{bmatrix}$ & $\begin{bmatrix}
						-0.000149          \\
						-0.001043\\
					\end{bmatrix}$
					&
					$\begin{bmatrix}
						(-0.015965 ,\, 0.016571)\\
						(-0.024911  ,\, 0.020223)
					\end{bmatrix}$\\\hline
					\multirow{2}{*}{10} & $\begin{bmatrix}
						0.012425 \\ -0.023567
					\end{bmatrix}$ & $\begin{bmatrix}
						-0.000223       \\
						-0.000900
					\end{bmatrix}$
					&
					$\begin{bmatrix}
						(-0.015342 ,\, 0.016601)\\
						(-0.023752 ,\, 0.023078)
					\end{bmatrix}$
			\end{tabular}}
		\end{table}

	\clearpage
	\section{Conclusions}
	\label{sec6}
	
This paper proposes an estimation, inference, and forecasting methodology for multivariate threshold autoregressive models (MTARs). The methodology is based on the Bayesian approach using MCMC methods. The steps are the following: ($i$) identification of the structural parameters and the distribution of the noise process, ($ii$) joint estimation of the non-structural parameters and the extra parameter associated with the noise process distribution, and ($iii$) forecasting from the fitted model via predictive distribution. Simulation experiments were carried out to assess the performance of the proposed methodology, in which prediction intervals were calculated to determine how often they captured the real values. Additionally, the percentage of times credible intervals capture true parameter values, and the empirical biases for thresholds and extra parameters of noise process distributions were calculated, indicating good estimation and forecasting performance. In general, for a fixed sample size, the methodology performs worse when the number of regimes is increased. Furthermore, the WAIC and DIC criteria can be used to select between a linear VAR and MTAR model, as well as to determine the number of regimes and autoregressive orders. Future work needs to provide tools to check the assumptions made for the noise process and propose a method for dealing with asymmetric noise distributions.

	\appendix
	
	\section{Appendix Estimation}
	\label{appendix:A}
	
	
\begin{table}[!ht]
	
	\centering
	\caption{Percentage of times that the true parameter values lie on the $95\%$ credible intervals considering $M_1$ with sample size $300$. For the delay parameter $h$ is considered the percentage of times that posterior mode coincides with the true delay value.}\label{CIM1300}
	\addtolength{\tabcolsep}{-5pt}
\resizebox{\columnwidth}{!}{
		\begin{tabular}{c|c|c|c|c|c|c}\hline       
			&  \multicolumn{6}{c}{Distribution of the noise process} \\\cline{2-7}
			& \multirow{2}{*}{Gaussian}&  \multirow{2}{*}{Student-$t$($\nu\!=\!3$)} & \multirow{2}{*}{Slash($\nu\!=\!6$)} & Contaminated & Symmetric & \multirow{2}{*}{Laplace} \\ 
			& & &  & normal      & hyperbolic& \\
			& & &  & ($\nu_1\!=\!0.05,\nu_2\!=\!0.1$)& ($\nu\!=\!0.11$)          & \\\hline
			Regime 1 & & & & & & \\
			$\phi_0^{(1)}$ & $\begin{bmatrix}
				93.5\\
				95.6\\
				96.1
			\end{bmatrix}$ &$\begin{bmatrix}
				93.3\\
				93.6\\
				93.0\\
			\end{bmatrix}$ &$\begin{bmatrix}
				94.3\\
				93.7\\
				93.2
			\end{bmatrix}$ & $\begin{bmatrix}
				92.2\\
				93.8\\
				95.2
			\end{bmatrix}$ & $\begin{bmatrix}
				93.7\\
				94.3\\
				95.1
			\end{bmatrix}$ & $\begin{bmatrix}
				94.2\\
				95.2\\
				93.7\\
			\end{bmatrix}$\\ \hline
			
			$\bm{\phi}_1^{(1)}$      & $\begin{bmatrix}
				95.2& 95.0& 94.6\\
				93.9& 93.2& 95.8\\
				95.4& 94.3& 96.0
			\end{bmatrix}$ & $\begin{bmatrix}
				93.4& 93.7& 95.0\\
				93.0& 93.5& 93.4\\
				94.1& 94.2& 94.7
			\end{bmatrix}$ & $\begin{bmatrix}
				93.5& 93.9& 92.9\\
				93.9& 94.1& 93.1\\
				92.9& 93.6& 92.0
			\end{bmatrix}$& $\begin{bmatrix}
				94.7& 93.9& 93.1\\
				94.1& 93.6& 94.1\\
				94.3& 93.6& 95.2
			\end{bmatrix}$ & $\begin{bmatrix}
				93.2& 94.5& 93.7\\
				94.3& 94.3& 93.4\\
				94.2& 94.7& 96.1
			\end{bmatrix}$ & $\begin{bmatrix}
				94.5& 93.0& 94.0\\
				94.0& 95.2& 93.7\\
				94.1& 93.1& 93.9
			\end{bmatrix}$\\ \hline
			$\bm{\beta}_1^{(1)}$      & $\begin{bmatrix}
				93.4& 95.1\\
				93.8& 95.2\\
				94.5& 95.3
			\end{bmatrix}$ & $\begin{bmatrix}
				92.8& 94.1\\
				92.9& 93.0\\
				93.5& 94.3
			\end{bmatrix}$ & $\begin{bmatrix}
				92.9& 94.6\\
				94.2& 94.8\\
				92.2& 94.0
			\end{bmatrix}$& $\begin{bmatrix}
				92.1& 93.5\\
				94.0& 92.3\\
				95.3& 93.3
			\end{bmatrix}$ & $\begin{bmatrix}
				92.8& 94.0\\
				94.5& 93.1\\
				94.3& 93.7
			\end{bmatrix}$ & $\begin{bmatrix}
				93.6& 94.7\\
				94.4& 93.3\\
				93.2& 93.4
			\end{bmatrix}$
			\\\hline
			$\bm{\Sigma}^{(1)}$  & $\begin{bmatrix}
				94.5& 95.0& 95.8\\
				95.0& 93.4& 94.6\\
				95.8& 94.6& 93.2
			\end{bmatrix}$ & $\begin{bmatrix}
				70.2& 94.8& 95.2\\
				94.8& 72.4& 96.1\\
				95.2& 96.1& 70.7
			\end{bmatrix}$ &$\begin{bmatrix}
				50.7& 96.2& 95.5\\
				96.2& 50.8& 95.4\\
				95.5& 95.4& 52.4
			\end{bmatrix}$ & $\begin{bmatrix}
				81.2& 93.1& 94.6\\
				93.1& 79.7& 93.7\\
				94.6& 93.7& 79.0
			\end{bmatrix}$ & $\begin{bmatrix}
				0.768& 0.941& 0.947\\
				0.941& 0.781& 0.958\\
				0.947& 0.958& 0.765
			\end{bmatrix}$ & $\begin{bmatrix}
				0.771& 0.951& 0.947\\
				0.951& 0.791& 0.955\\
				0.947& 0.955& 0.795
			\end{bmatrix}$\\ \hline
			Regime 2 & & & & & & \\
			$\phi_0^{(2)}$ & $\begin{bmatrix}
				94.1\\
				94.9\\
				94.3\\
			\end{bmatrix}$ & $\begin{bmatrix}
				91.1\\
				94.3\\
				93.5
			\end{bmatrix}$ & $\begin{bmatrix}
				92.8\\
				93.7\\
				92.2
			\end{bmatrix}$ & $\begin{bmatrix}
				92.7\\
				93.3\\
				94.7
			\end{bmatrix}$ & $\begin{bmatrix}
				93.8\\
				93.6\\
				93.2
			\end{bmatrix}$ & $\begin{bmatrix}
				0.936\\
				0.945\\
				0.940
			\end{bmatrix}$\\ \hline
			
			$\bm{\phi}_1^{(2)}$      & $\begin{bmatrix}
				94.6& 95.6& 95.6\\
				95.6& 94.9& 95.6\\
				94.3& 94.9& 95.0
			\end{bmatrix}$ & $\begin{bmatrix}
				92.8& 92.8& 93.2\\
				92.7& 93.8& 94.2\\
				94.5& 94.1& 95.6\\
			\end{bmatrix}$ & $\begin{bmatrix}
				93.7& 92.9& 92.8\\
				92.2& 93.0& 94.5\\
				93.0& 92.5& 92.9
			\end{bmatrix}$ & $\begin{bmatrix}
				93.2& 93.5& 93.6\\
				93.5& 94.8& 94.7\\
				92.8& 93.1& 93.8\\
			\end{bmatrix}$ & $\begin{bmatrix}
				945.& 94.7& 94.8\\
				93.0& 94.3& 94.5\\
				93.7& 92.9& 94.2
			\end{bmatrix}$ & $\begin{bmatrix}
				93.8& 95.2& 94.4\\
				93.6& 94.4& 93.6\\
				93.8& 93.0& 93.3
			\end{bmatrix}$\\ \hline
			
			$\bm{\phi}_2^{(2)}$      & $\begin{bmatrix}
				94.7& 94.7& 94.0\\
				94.8& 94.9& 94.7\\
				0.935& 0.943& 0.958
			\end{bmatrix}$ & $\begin{bmatrix}
				92.8& 93.7& 92.5\\
				93.9& 93.7& 93.8\\
				0.938& 0.946& 0.929
			\end{bmatrix}$ & $\begin{bmatrix}
				92.7& 92.3& 93.5\\
				92.8& 94.4& 94.1\\
				92.3& 92.7& 93.3
			\end{bmatrix}$ & $\begin{bmatrix}
				93.8& 93.3& 94.0\\
				94.3& 94.6& 93.3\\
				92.9& 94.3& 93.5
			\end{bmatrix}$ & $\begin{bmatrix}
				93.4& 94.3& 95.9\\
				95.2& 93.9& 95.1\\
				94.4& 92.6& 94.2
			\end{bmatrix}$ & $\begin{bmatrix}
				94.1& 94.2& 94.5\\
				94.6& 94.2& 94.2\\
				92.1& 93.6& 93.0
			\end{bmatrix}$\\ \hline

			$\bm{\Sigma}^{(2)}$  & $\begin{bmatrix}
				94.7& 95.2& 96.6\\
				95.2& 93.4& 95.7\\
				96.6& 95.7& 94.1
			\end{bmatrix}$ & $\begin{bmatrix}
				68.7& 94.9& 96.2\\
				94.9& 65.3& 95.8\\
				96.2& 95.8& 68.3
			\end{bmatrix}$ & $\begin{bmatrix}
				48.9& 96.3& 95.4\\
				96.3& 46.4& 94.2\\
				95.4& 94.2& 48.5
			\end{bmatrix}$ & $\begin{bmatrix}
				80.8& 94.0& 95.5\\
				94.0& 79.6& 94.2\\
				95.5& 94.2&77.8\\
			\end{bmatrix}$ & $\begin{bmatrix}
				75.2& 96.1& 95.3\\
				96.1& 72.5& 93.1\\
				95.3& 93.1& 72.8
			\end{bmatrix}$ & $\begin{bmatrix}
				78.0& 95.0& 95.0\\
				95.0& 76.0& 94.6\\
				95.0& 94.6& 74.8
			\end{bmatrix}$\\ \hline
			$c$ & 78.3  & 75.3 &61.1 & 62.7 & 63.2 & 63.5 
			\\
			$h$ & 100  & 100 & 100 & 100 &100& 100 \\
			$\nu$ & &59.3 & 23.2& $\begin{bmatrix}
				84.3\\ 88.1
			\end{bmatrix}$ & 100 &  \\\hline
	\end{tabular} }
\end{table}

\begin{table}[!ht]
	\centering
	\caption{Bias for the threshold vector ($c$) and the relative bias$\times100$ for extra parameter ($\nu$) considering $M_1$ with sample size $300$.}
	\label{BiasM1300}
\resizebox{\columnwidth}{!}{
		\begin{tabular}{c|c|c|c|c|c|c}\hline       
			&  \multicolumn{6}{c}{Distribution of the noise process} \\\cline{2-7}
			& \multirow{2}{*}{Gaussian}&  \multirow{2}{*}{Student-$t$($3$)} & \multirow{2}{*}{Slash($6$)} & Contaminated & Symmetric & \multirow{2}{*}{Laplace} \\ 
			& & &  & normal      & hyperbolic& \\
			& & &  & ($\nu_1\!=\!0.05,\nu_2\!=\!0.1$)& ($\nu\!=\!0.11$)          & \\\hline
			$c$ & -0.00102386 & 0.00026258 & -0.00017922 & 0.00033395 & $-0.00164022$ & -0.00084028\\
			$\nu$ & & 7.596  & 34.263   & $\begin{bmatrix}
				189.268\\ 74.614
			\end{bmatrix}$ & 1.452  &  \\
	\end{tabular}}
\end{table}

\begin{table}[!ht]
	\centering
	\caption{Percentage of times that the true parameter values lie on the $95\%$ credible intervals considering $M_2$ with sample size $300$. For the delay parameter $h$ is considered the percentage of times that posterior mode coincides with the true delay value}\label{CIM2300}
	\resizebox{\columnwidth}{!}{
		\begin{tabular}{c|c|c|c|c|c|c}\hline       
			&  \multicolumn{6}{c}{Distribution} \\\cline{2-7}
			& \multirow{2}{*}{Gaussian}&  \multirow{2}{*}{Student-t} & \multirow{2}{*}{Slash} & Contaminated & Symmetric & \multirow{2}{*}{Laplace} \\ 
			& & &  & normal& hyperbolic& \\\hline
			Regime 1 & & & & & & \\
			$\phi_0^{(1)}$ & $\begin{bmatrix}
				92.3\\
				89.2
			\end{bmatrix}$ &$\begin{bmatrix}
				91.4\\
				91.6
			\end{bmatrix}$ &$\begin{bmatrix}
				90.4\\
				91.6
			\end{bmatrix}$ & $\begin{bmatrix}
				92.7\\
				93.7
			\end{bmatrix}$ & $\begin{bmatrix}
				92.6\\
				92.9
			\end{bmatrix}$ & $\begin{bmatrix}
				93.5\\
				91.3
			\end{bmatrix}$\\ \hline
			
			$\bm{\phi}_1^{(1)}$      & $\begin{bmatrix}
				92.2 &95.1\\
				91.8 &91.1
			\end{bmatrix}$ & $\begin{bmatrix}
				91.6& 93.3\\
				92.1& 90.6\\
			\end{bmatrix}$ & $\begin{bmatrix}
				90.6& 93.6\\
				91.8& 92.1
			\end{bmatrix}$& $\begin{bmatrix}
				91.7& 94.7\\
				92.8& 91.8\\
			\end{bmatrix}$ & $\begin{bmatrix}
				90.2& 92.5\\
				93.0& 87.2
			\end{bmatrix}$ & $\begin{bmatrix}
				90.8 &93.9\\
				93.1 &91.9
			\end{bmatrix}$\\ \hline
			$\bm{\Sigma}^{(1)}$  & $\begin{bmatrix}
				81.3 &91.5\\
				91.5 &88.0
			\end{bmatrix}$ & $\begin{bmatrix}
				81.5& 95.2\\
				95.2& 82.8\\
			\end{bmatrix}$ &$\begin{bmatrix}
				74.0& 95.1\\
				95.1& 71.3
			\end{bmatrix}$ & $\begin{bmatrix}
				53.9& 91.5\\
				91.5& 51.0\\
			\end{bmatrix}$ & $\begin{bmatrix}
				77.9 &94.4\\
				94.4 &77.8
			\end{bmatrix}$ & $\begin{bmatrix}
				90.9& 95.4\\
				95.4& 89.2
			\end{bmatrix}$\\ \hline
			Regime 2 & & & & & & \\
			$\phi_0^{(2)}$ & $\begin{bmatrix}
				92.8\\
				90.9
			\end{bmatrix}$ & $\begin{bmatrix}
				92.1\\
				93.7\\
			\end{bmatrix}$ & $\begin{bmatrix}
				90.3\\
				90.8
			\end{bmatrix}$ & $\begin{bmatrix}
				93.9\\
				93.7
			\end{bmatrix}$ & $\begin{bmatrix}
				93.7\\
				94.4
			\end{bmatrix}$ & $\begin{bmatrix}
				92.5\\
				92.2
			\end{bmatrix}$\\ \hline
			
			$\bm{\phi}_1^{(2)}$      & $\begin{bmatrix}
				91.5 &95.5\\
				93.6 &89.2
			\end{bmatrix}$ & $\begin{bmatrix}
				94.2& 94.9\\
				93.6& 93.7\\
			\end{bmatrix}$ & $\begin{bmatrix}
				91.3 &91.0\\
				92.7 &92.2
			\end{bmatrix}$ & $\begin{bmatrix}
				92.9& 94.2\\
				93.5& 94.4
			\end{bmatrix}$ & $\begin{bmatrix}
				88.5& 94.5\\
				92.5& 88.1
			\end{bmatrix}$ & $\begin{bmatrix}
				91.1 &93.6\\
				93.4 &90.0
			\end{bmatrix}$\\ \hline
			$\bm{\Sigma}^{(2)}$  & $\begin{bmatrix}
				76.2 &85.2\\
				85.2 &75.0
			\end{bmatrix}$ & $\begin{bmatrix}
				90.4& 94.0\\
				94.0& 87.9\\
			\end{bmatrix}$ & $\begin{bmatrix}
				78. &94.0\\
				94.0 &79.1
			\end{bmatrix}$ & $\begin{bmatrix}
				55.0& 90.9\\
				90.9& 54.3
			\end{bmatrix}$ & $\begin{bmatrix}
				82.6 &94.8\\
				94.8 &81.6
			\end{bmatrix}$ & $\begin{bmatrix}
				89.6 &93.8\\
				93.8 &83.0
			\end{bmatrix}$\\ \hline
			Regime 3 & & & & & & \\
			$\phi_0^{(3)}$ & $\begin{bmatrix}
				93.4\\
				93.9
			\end{bmatrix}$ & $\begin{bmatrix}
				91.0\\
				94.6\\
			\end{bmatrix}$ &$\begin{bmatrix}
				91.4\\
				91.9
			\end{bmatrix}$ & $\begin{bmatrix}
				94.1\\
				93.4\\
			\end{bmatrix}$ & $\begin{bmatrix}
				91.7\\
				93.2
			\end{bmatrix}$ & $\begin{bmatrix}
				91.9\\
				92.8
			\end{bmatrix}$\\ \hline
			
			$\bm{\phi}_1^{(3)}$      & $\begin{bmatrix}
				95.1 &94.6\\
				95.1 &90.2
			\end{bmatrix}$ & $\begin{bmatrix}
				94.6& 93.9\\
				93.7& 92.7\\
			\end{bmatrix}$ & $\begin{bmatrix}
				93.3 &93.5\\
				93.5 &92.2
			\end{bmatrix}$ & $\begin{bmatrix}
				94.3& 93.0\\
				94.2& 92.0\\
			\end{bmatrix}$ & $\begin{bmatrix}
				91.9 & 93.6\\
				91.8 &90.4
			\end{bmatrix}$ & $\begin{bmatrix}
				94.6 &93.0\\
				94.1 &90.3
			\end{bmatrix}$\\ \hline
			$\bm{\Sigma}^{(3)}$  & $\begin{bmatrix}
				89.2 &88.9\\
				88.9 &85.0
			\end{bmatrix}$  & $\begin{bmatrix}
				84.3& 94.1\\
				94.1& 85.4
			\end{bmatrix}$ &$\begin{bmatrix}
				72.0 &95.0\\
				95.0 &72.7
			\end{bmatrix}$ & $\begin{bmatrix}
				52.1& 90.6\\
				90.6& 52.4
			\end{bmatrix}$ & $\begin{bmatrix}
				75.8& 95.6\\
				95.6& 77.9
			\end{bmatrix}$ & $\begin{bmatrix}
				89.8 &96.3\\
				96.3 &86.0
			\end{bmatrix}$\\ \hline
			$\begin{matrix}
				c_1\\
				c_2
			\end{matrix}$ & $\begin{bmatrix}
				17.0\\
				17.1
			\end{bmatrix}$ & $\begin{bmatrix}
				14.6\\
				14.4
			\end{bmatrix}$ &$\begin{bmatrix}
				13.3\\
				11.9
			\end{bmatrix}$ & $\begin{bmatrix}
				11.2\\
				12.0
			\end{bmatrix}$ & $\begin{bmatrix}
				17.0\\
				17.1
			\end{bmatrix}$ & $\begin{bmatrix}
				17.0\\
				17.1
			\end{bmatrix}$ 
			\\
			$h$ & 99.9  & 100 & 98.8 & 99.8 &94.9 & 97.1 \\
			$\nu$ & &56.0 & 48.6& $\begin{bmatrix}
				51.7\\
				47.2
			\end{bmatrix}$ & 71.4 &  \\\hline
	\end{tabular} }
\end{table}

\begin{table}[!ht]
	\centering
	\caption{Relative bias for the threshold vector ($c$) and the extra parameter ($\nu$) considering $M_2$ with sample size $300$.}
	\label{BiasM2300}
	\resizebox{\columnwidth}{!}{
		\begin{tabular}{c|c|c|c|c|c|c}\hline       
			&  \multicolumn{6}{c}{Distribution} \\\cline{2-7}
			& \multirow{2}{*}{Gaussian}&  \multirow{2}{*}{Student-$t$($\nu\!=\!5$)} & \multirow{2}{*}{Slash($\nu\!=\!4$)} & Contaminated & Symmetric & \multirow{2}{*}{Laplace} \\ 
			& & &  & normal& hyperbolic& \\
			& & &  & ($\nu_1\!=\!0.08,\nu_2\!=\!0.012$)& ($\nu\!=\!0.12$)& \\\hline
			$c$ & $\begin{bmatrix}
				1.010\\
				0.001
			\end{bmatrix}$ & $\begin{bmatrix}
				0.626\\
				0.0805
			\end{bmatrix}$ & $\begin{bmatrix}
				0.135\\
				0.399
			\end{bmatrix}$ & $\begin{bmatrix}
				0.288\\
				0.010
			\end{bmatrix}$ & $\begin{bmatrix}
				0.474\\
				0.702
			\end{bmatrix}$ & $\begin{bmatrix}
				0.093\\
				0.268
			\end{bmatrix}$\\
			$\nu$ &  & 22.006 & 8.851   & $\begin{bmatrix}
				417.3121 \\
				1668.943
			\end{bmatrix}$ & 7.057  &  \\
	\end{tabular}}
\end{table}

\newpage
	\section{Appendix Forecasting}
	\label{appendix:B}

\begin{table}[!ht]
	\centering
	\caption{Percentage of times that the true values lie on $95\%$ credible intervals considering $M_1$ with sample size $300$.}\label{FORECIM1300}
\resizebox{\columnwidth}{!}{
		\begin{tabular}{c|c|c|c|c|c|c}\hline       
			&  \multicolumn{6}{c}{Distribution} \\\cline{2-7}
			step ahead& \multirow{2}{*}{Gaussian}&  \multirow{2}{*}{Student-t} & \multirow{2}{*}{Slash} & Contaminated & Symmetric & \multirow{2}{*}{Laplace} \\ 
			& & &  & normal& hyperbolic& \\\hline
			$\bm{y}_{t+1}$ & $\begin{bmatrix}
				95.1\\ 94.1\\ 95.2
			\end{bmatrix}$ &$\begin{bmatrix}
				94.9\\  95.7\\  94.3
			\end{bmatrix}$ &$\begin{bmatrix}
				94.3\\  94.8\\  92.7
			\end{bmatrix}$ & $\begin{bmatrix}
				94.8\\  95.1\\  94.6
			\end{bmatrix}$ & $\begin{bmatrix}
				92.8\\  92.7\\  93.3
			\end{bmatrix}$ & $\begin{bmatrix}
				93.6\\  93.3\\  93.8
			\end{bmatrix}$\\ \hline
			$\bm{y}_{t+2}$ & $\begin{bmatrix}
				94.5\\  94.6\\  94.2
			\end{bmatrix}$ &$\begin{bmatrix}
				95.6\\  96.1\\  94.6
			\end{bmatrix}$ &$\begin{bmatrix}
				94.3\\  94.6\\  94.4
			\end{bmatrix}$ & $\begin{bmatrix}
				94.3\\  95.1\\  94.4
			\end{bmatrix}$ & $\begin{bmatrix}
				93.1\\  93.1\\  93.6
			\end{bmatrix}$ & $\begin{bmatrix}
				94.0\\  91.6\\  93.6
			\end{bmatrix}$\\ \hline
			$\bm{y}_{t+3}$ & $\begin{bmatrix}
				94.4\\  96.1\\  941
			\end{bmatrix}$ &$\begin{bmatrix}
				95.5\\  94.5\\  94.6
			\end{bmatrix}$ &$\begin{bmatrix}
				95.1\\  94.7\\  94.4
			\end{bmatrix}$ & $\begin{bmatrix}
				94.9\\  93.7\\  94.7
			\end{bmatrix}$ & $\begin{bmatrix}
				92.8\\  92.6\\  93.1
			\end{bmatrix}$ & $\begin{bmatrix}
				94.1\\  91.5\\  93.2
			\end{bmatrix}$\\ \hline
			$\bm{y}_{t+4}$ & $\begin{bmatrix}
				94.9\\  92.8\\  95.1
			\end{bmatrix}$ &$\begin{bmatrix}
				96.1\\  96.0\\  95.2
			\end{bmatrix}$ &$\begin{bmatrix}
				94.9\\  95.6\\  95.5
			\end{bmatrix}$ & $\begin{bmatrix}
				94.2\\  93.9\\  94.5
			\end{bmatrix}$ & $\begin{bmatrix}
				93.2\\  93.0\\  93.6
			\end{bmatrix}$ & $\begin{bmatrix}
				93.6\\  93.6\\  91.4
			\end{bmatrix}$\\ \hline
			$\bm{y}_{t+5}$ & $\begin{bmatrix}
				95.6\\  94.7\\  94.1
			\end{bmatrix}$ &$\begin{bmatrix}
				94.9\\  95.8\\  95.4
			\end{bmatrix}$ &$\begin{bmatrix}
				95.5\\  95.5\\  95.1
			\end{bmatrix}$ & $\begin{bmatrix}
				94.6\\  93.8\\  94.1
			\end{bmatrix}$ & $\begin{bmatrix}
				93.5\\  93.6\\  93.4
			\end{bmatrix}$ & $\begin{bmatrix}
				93.9\\  93.4\\  94.3
			\end{bmatrix}$\\ \hline
			$\bm{y}_{t+6}$ & $\begin{bmatrix}
				94.6\\  94.2\\  946.
			\end{bmatrix}$ &$\begin{bmatrix}
				95.2\\  95.1\\  96.7
			\end{bmatrix}$ &$\begin{bmatrix}
				95.3\\  95.9\\  95.5
			\end{bmatrix}$ & $\begin{bmatrix}
				94.4\\  93.4\\  94.9
			\end{bmatrix}$ & $\begin{bmatrix}
				93.6\\  93.7\\  93.8
			\end{bmatrix}$ & $\begin{bmatrix}
				94.3\\  92.8\\  92.7
			\end{bmatrix}$\\ \hline
			$\bm{y}_{t+7}$ & $\begin{bmatrix}
				94.6\\  94.5\\  94.1
			\end{bmatrix}$ &$\begin{bmatrix}
				95.6\\  95.1\\  95.9
			\end{bmatrix}$ &$\begin{bmatrix}
				94.8\\  95.3\\  95.5
			\end{bmatrix}$ & $\begin{bmatrix}
				95.2\\  94.7\\  94.3
			\end{bmatrix}$ & $\begin{bmatrix}
				93.4\\  93.1\\  94.2
			\end{bmatrix}$ & $\begin{bmatrix}
				92.9\\  93.4\\  93.1
			\end{bmatrix}$\\ \hline
			$\bm{y}_{t+8}$ & $\begin{bmatrix}
				95.2\\  94.9\\  94.3
			\end{bmatrix}$ &$\begin{bmatrix}
				95.4\\  96.6\\  95.6
			\end{bmatrix}$ &$\begin{bmatrix}
				96.2\\  95.5\\  95.5
			\end{bmatrix}$ & $\begin{bmatrix}
				95.8\\  94.4\\  95.5
			\end{bmatrix}$ & $\begin{bmatrix}
				94.3\\  93.9\\  94.0
			\end{bmatrix}$ & $\begin{bmatrix}
				92.6\\  92.1\\  93.6
			\end{bmatrix}$\\ \hline
			$\bm{y}_{t+9}$ & $\begin{bmatrix}
				95.4\\  94.6\\  93.3
			\end{bmatrix}$ &$\begin{bmatrix}
				95.9\\  96.3\\  95.6
			\end{bmatrix}$ &$\begin{bmatrix}
				95.8\\  95.8\\  94.8
			\end{bmatrix}$ & $\begin{bmatrix}
				95.0\\  93.9\\  93.3\\
			\end{bmatrix}$ & $\begin{bmatrix}
				93.8\\  92.4\\  91.4
			\end{bmatrix}$ & $\begin{bmatrix}
				92.4\\  92.4\\  92.1
			\end{bmatrix}$\\ \hline
			$\bm{y}_{t+10}$ & $\begin{bmatrix}
				95.8\\  93.8\\  93.7
			\end{bmatrix}$ &$\begin{bmatrix}
				96.4\\  96.8\\  95.4
			\end{bmatrix}$ &$\begin{bmatrix}
				96.0\\  96.1\\  94.9
			\end{bmatrix}$ & $\begin{bmatrix}
				95.3\\  94.4\\  95.1
			\end{bmatrix}$ & $\begin{bmatrix}
				93.8\\  93.9\\  93.3
			\end{bmatrix}$ & $\begin{bmatrix}
				91.9\\  91.7\\  93.5
			\end{bmatrix}$\\ \hline
	\end{tabular} }
\end{table}

\begin{table}[!ht]
	\centering
	\caption{Percentage of times that the true values lie on $95\%$ credible intervals considering $M_2$ with sample size $300$.}\label{FORECIM2300}
\resizebox{\columnwidth}{!}{
		\begin{tabular}{c|c|c|c|c|c|c}\hline       
			&  \multicolumn{6}{c}{Distribution} \\\cline{2-7}
			step ahead& \multirow{2}{*}{Gaussian}&  \multirow{2}{*}{Student-t} & \multirow{2}{*}{Slash} & Contaminated & Symmetric & \multirow{2}{*}{Laplace} \\ 
			& & &  & normal& hyperbolic& \\\hline
			$\bm{y}_{t+1}$ & $\begin{bmatrix}
				95.6\\  95.2
			\end{bmatrix}$ &$\begin{bmatrix}
				94.9\\  93.9
			\end{bmatrix}$ &$\begin{bmatrix}
				96.8\\  96.1
			\end{bmatrix}$ & $\begin{bmatrix}
				95.1\\  94.1
			\end{bmatrix}$ & $\begin{bmatrix}
				92.7\\  92.4
			\end{bmatrix}$ & $\begin{bmatrix}
				93.5\\  93.2
			\end{bmatrix}$\\ \hline
			$\bm{y}_{t+2}$ & $\begin{bmatrix}
				95.3\\  95.4
			\end{bmatrix}$ &$\begin{bmatrix}
				94.5\\  92.8
			\end{bmatrix}$ &$\begin{bmatrix}
				96.2\\  96.1
			\end{bmatrix}$ & $\begin{bmatrix}
				95.4\\  94.3
			\end{bmatrix}$ & $\begin{bmatrix}
				92.9\\  95.7
			\end{bmatrix}$ & $\begin{bmatrix}
				95.0\\  93.6
			\end{bmatrix}$\\ \hline
			$\bm{y}_{t+3}$ & $\begin{bmatrix}
				94.9\\  94.8
			\end{bmatrix}$ &$\begin{bmatrix}
				94.8\\  95.5
			\end{bmatrix}$ &$\begin{bmatrix}
				96.6\\  95.4
			\end{bmatrix}$ & $\begin{bmatrix}
				94.4\\  93.7
			\end{bmatrix}$ & $\begin{bmatrix}
				92.8\\  94.8
			\end{bmatrix}$ & $\begin{bmatrix}
				94.1\\  94.2
			\end{bmatrix}$\\ \hline
			$\bm{y}_{t+4}$ & $\begin{bmatrix}
				95.9\\  94.4
			\end{bmatrix}$ &$\begin{bmatrix}
				96.1\\  96.1
			\end{bmatrix}$ &$\begin{bmatrix}
				96.5\\  96.5
			\end{bmatrix}$ & $\begin{bmatrix}
				94.0\\  94.5
			\end{bmatrix}$ & $\begin{bmatrix}
				93.7\\  94.6
			\end{bmatrix}$ & $\begin{bmatrix}
				94.6\\  95.1
			\end{bmatrix}$\\ \hline
			$\bm{y}_{t+5}$ & $\begin{bmatrix}
				95.5\\  94.7
			\end{bmatrix}$ &$\begin{bmatrix}
				94.6\\  95.2
			\end{bmatrix}$ &$\begin{bmatrix}
				96.2\\  97.1
			\end{bmatrix}$ & $\begin{bmatrix}
				93.9\\  93.7
			\end{bmatrix}$ & $\begin{bmatrix}
				94.3\\  93.1
			\end{bmatrix}$ & $\begin{bmatrix}
				94.4\\  95.0
			\end{bmatrix}$\\ \hline
			$\bm{y}_{t+6}$ & $\begin{bmatrix}
				94.8\\  96.2
			\end{bmatrix}$ &$\begin{bmatrix}
				94.1\\  94.2
			\end{bmatrix}$ &$\begin{bmatrix}
				96.5\\  96.9
			\end{bmatrix}$ & $\begin{bmatrix}
				94.0\\  94.8
			\end{bmatrix}$ & $\begin{bmatrix}
				92.8\\  93.8
			\end{bmatrix}$ & $\begin{bmatrix}
				93.6\\  94.1
			\end{bmatrix}$\\ \hline
			$\bm{y}_{t+7}$ & $\begin{bmatrix}
				94.7\\  95.2
			\end{bmatrix}$ &$\begin{bmatrix}
				94.9\\  94.9
			\end{bmatrix}$ &$\begin{bmatrix}
				96.8\\  96.6
			\end{bmatrix}$ & $\begin{bmatrix}
				94.4\\  93.9
			\end{bmatrix}$ & $\begin{bmatrix}
				92.6\\  93.9
			\end{bmatrix}$ & $\begin{bmatrix}
				94.2\\  94.5
			\end{bmatrix}$\\ \hline
			$\bm{y}_{t+8}$ & $\begin{bmatrix}
				94.8\\  96.1
			\end{bmatrix}$ &$\begin{bmatrix}
				95.2\\  94.4
			\end{bmatrix}$ &$\begin{bmatrix}
				96.6\\  96.9
			\end{bmatrix}$ & $\begin{bmatrix}
				94.8\\  95.6
			\end{bmatrix}$ & $\begin{bmatrix}
				94.6\\  94.4
			\end{bmatrix}$ & $\begin{bmatrix}
				94.6\\  93.0
			\end{bmatrix}$\\ \hline
			$\bm{y}_{t+9}$ & $\begin{bmatrix}
				95.2\\  94.7
			\end{bmatrix}$ &$\begin{bmatrix}
				95.3\\  95.2
			\end{bmatrix}$ &$\begin{bmatrix}
				96.4\\  96.0
			\end{bmatrix}$ & $\begin{bmatrix}
				94.8\\  95.7
			\end{bmatrix}$ & $\begin{bmatrix}
				93.7\\  94.1
			\end{bmatrix}$ & $\begin{bmatrix}
				95.5\\  93.4
			\end{bmatrix}$\\ \hline
			$\bm{y}_{t+10}$ & $\begin{bmatrix}
				95.9\\  93.4
			\end{bmatrix}$ &$\begin{bmatrix}
				95.0\\  95.7
			\end{bmatrix}$ &$\begin{bmatrix}
				96.2\\  96.2
			\end{bmatrix}$ & $\begin{bmatrix}
				94.9\\  94.8
			\end{bmatrix}$ & $\begin{bmatrix}
				93.4\\  93.6
			\end{bmatrix}$ & $\begin{bmatrix}
				93.3\\  93.9
			\end{bmatrix}$\\ \hline
	\end{tabular} }
\end{table}

\clearpage

\bibliographystyle{plain}

\end{document}